\documentclass[5p]{elsarticle}

\usepackage{lineno}

\usepackage{latexsym}
\usepackage{color}
\usepackage{bm}
\usepackage{enumitem}
\usepackage{array}
\usepackage[english]{babel}
\usepackage{tensor}
\usepackage{float}
\usepackage{ulem}
\usepackage[utf8]{inputenc}
\usepackage[T1]{fontenc}
\usepackage{comment}

\begin{document}

\begin{frontmatter}

\title{ A designer approach to $f(Q)$ gravity and cosmological implications}

\author{In\^es S.~Albuquerque}
\ead{isalbuquerque@fc.ul.pt}
\author{Noemi Frusciante}
\ead{nfrusciante@fc.ul.pt}
\address{\smallskip
Instituto de Astrof\'isica e Ci\^encias do Espa\c{c}o, Faculdade de Ci\^encias da Universidade de Lisboa, Edificio C8, Campo Grande, P-1749016, Lisboa, Portugal 
}

\begin{abstract}
We investigate the evolution of linear perturbations in the Symmetric Teleparallel Gravity, namely $f(Q)$ gravity, for which we design the $f(Q)$ function to match specific expansion histories. We consider different evolutions of the effective dark energy equation of state, $w_Q(a)$, which includes $w_Q=-1$, a constant $w_Q \neq -1$ and a fast varying equation of state. We identify clear patterns in the effective gravitational coupling, which accordingly modifies the linear growth of large scale structures. We provide theoretical predictions for the product of the growth rate $\tilde{f}$ and the root mean square of matter fluctuations $\sigma_8$, namely $\tilde{f}\sigma_8$ and for the sign of the cross-correlation power spectrum of the galaxy fluctuations and the cosmic microwave background radiation anisotropies. These properties can be used to distinguish the $f(Q)$ gravity from the standard cosmological model using accurate cosmological observations.
\end{abstract}


\end{frontmatter}

\section{Introduction}

The observed late-time accelerated expansion of the Universe \cite{SupernovaSearchTeam:1998fmf,SupernovaCosmologyProject:1998vns,WMAP:2003elm,SDSS:2005xqv,SDSS:2014iwm,Planck:2015fie,Planck:2015bpv}, is modelled by the cosmological constant $\Lambda$ within the Einstein's theory of General Relativity (GR). The resulting cosmological model, known as $\Lambda$ Cold Dark Matter ($\Lambda$CDM), 
is recently facing some relevant challenges. In fact, besides the well known theoretical problems \cite{Weinberg:1988cp,Carroll:2000fy,Velten:2014nra,Joyce:2014kja,Padilla:2015aaa}, the $\Lambda$CDM model suffers from some mild observational tensions as well, namely on the measurements of the value of the Hubble constant $H_0$ \cite{Riess:2019cxk,Wong:2019kwg,BOSS:2014hwf,Dawson:2012va,SDSS:2008tqn,Freedman:2019jwv,DiValentino:2020zio} and the present-time amplitude of the matter power spectrum $\sigma_{8}^{0}$ \cite{deJong:2015wca,Hildebrandt:2016iqg,Kuijken:2015vca,DiValentino:2020vvd}. Alternatives beyond $\Lambda$CDM, among which those modifying the long range gravitational interaction known as Modified Gravity (MG) theories~\cite{Joyce:2014kja,Lue:2004rj,Copeland:2006wr,Silvestri:2009hh,Nojiri:2010wj,Tsujikawa:2010zza,Capozziello:2011et,Clifton:2011jh,Bamba:2012cp,Koyama:2015vza,Avelino:2016lpj,Joyce:2016vqv,Nojiri:2017ncd,Ferreira:2019xrr,Kobayashi:2019hrl,Frusciante:2019xia,CANTATA:2021ktz,Bahamonde:2021gfp}, have been deeply scrutinized, with some of them proving able to alleviate the tensions within $1-2\sigma$ \cite{Nunes:2018xbm,Zumalacarregui:2020cjh,Belgacem:2017cqo,Rossi:2019lgt,Peirone:2019aua,Frusciante:2019puu,Heisenberg:2020xak,Barros:2020bgg,Barros:2018efl}, see also \cite{DiValentino:2021izs} for a review.

$f(Q)$ gravity \cite{BeltranJimenez:2017tkd,Harko:2018gxr,Xu:2019sbp,Jarv:2018bgs,Runkla:2018xrv} is a MG theory which recently attracted a lot of attention.  It belongs to the Symmetric Teleparallel Gravity \cite{Nester:1998mp,Adak:2008gd,Adak:2018vzk} in which gravity is attributed to the non-metricity and where $f(Q)$ is a general function of the non-metricity scalar, $Q$.
The theory does not show strong coupling problems when considering perturbations around a Friedmann Lema{\^i}tre Robertson Walker (FLRW) background \cite{BeltranJimenez:2019tjy}. As such, the main linear perturbation equations for scalar, tensor and vector modes were  derived, showing  specific modifications with respect to the $\Lambda$CDM model \cite{BeltranJimenez:2019tjy}. 
These motivated further exploration of their impact on the cosmological observables. Constraints at background level have been provided using the expansion rate data from  early type galaxies, Supernovae type Ia (SNIa), quasars, gamma ray bursts, Baryonic Acoustic Oscillations (BAO) data, and Cosmic Microwave Background (CMB) distance priors,
for different parametrizations of $f(Q)$ as an explicit function of redshift $z$ \cite{Lazkoz:2019sjl}.
A similar investigation has been performed using a power-law term for $f(Q)$, namely $f(Q) = Q + \beta Q^n$ \cite{Ayuso:2020dcu}, for which cosmological solutions and the evolution of the growth index of matter perturbations have been investigated as well \cite{Khyllep:2021pcu}.
Alternatively, an exponential form for the $f(Q)$ function has been recently proposed, explicitly $f(Q) = Q \, e^{\lambda \frac{Q_0}{Q}}$, for which a statistical preference over $\Lambda$CDM has been found when the combination of cosmic chronometers, SNIa and BAO datasets is considered \cite{Anagnostopoulos:2021ydo}. 
Regarding linear perturbations in $f(Q)$ gravity, in particular for the $f(Q)$ model which mimics an exact $\Lambda$CDM expansion history, modifications in the evolution of matter density fields  were tested against redshift space distortions (RSD) data, showing the ability for the model to alleviate the $\sigma_8$ tension  \cite{Barros:2020bgg}, while measurable effects have been identified to characterize  the matter power spectrum, the lensing effect on the CMB angular power spectrum, CMB temperature anisotropies and the  Gravitational Waves (GWs) propagation \cite{Frusciante:2021sio}. A joint analysis of CMB, BAO, RSD, SNIa, Weak Lensing (WL) and Galaxy Clustering (GC) data, was able to strongly constrain the model's parameter and it was found that the model can actually challenge the $\Lambda$CDM scenario \cite{Atayde:2021ujc}. Besides these results, $f(Q)$ gravity has been the subject of a variety of studies in many different directions \cite{Dialektopoulos:2019mtr,Jimenez:2019ovq,Bajardi:2020fxh,Flathmann:2020zyj,Khyllep:2021pcu,DAmbrosio:2020nev,Solanki:2021qni,Zhao:2021zab,Bohmer:2021eoo}.

Previous works follow a  common procedure, namely they choose the form for $f(Q)$ \textit{a priori} and then determine the corresponding expansion history and linear perturbation dynamics. In this work we will opt for the reverse approach: we instead fix the expansion history and then solve the background equations for the corresponding form of $f(Q)$. The expansion history will be, in practice, selected by the choice of evolution for the equation of state parameter for an effective dark energy component associated with $Q$, $w_Q$. This is known as the \textit{designer} approach,  previously applied to $f(R)$ theory \cite{Song:0610532,Pogosian:0709}. One advantage of this approach is that it allows one to identify the impact of a background evolution, which differs from the $\Lambda$CDM, on some physical quantities of relevance at linear perturbation level and to evaluate whether the parameters characterizing the effective equation of state are degenerate with the one affecting the perturbations only. Using this approach, we then identify the general features characterizing linear perturbations which can later be employed to constrain the model. In particular, we identify the \textit{effective gravitational coupling} as the source of the modifications of the gravitational interaction at large scales. Therefore we consider its signatures on related physical quantities, such as the growth of matter perturbations and the cross-correlation power spectrum of the CMB anisotropies
with galaxy distribution.

This work is organized as follows. In Section~\ref{Sec:fQbasics} we briefly review the basis of the $f(Q)$ gravity formalism. The \textit{designer} approach is then detailed in Section~\ref{Sec:designer}, which includes a study of the initial conditions and the choices for the effective dark energy equation of state. Then, in Section~\ref{Sec:effectivecoupling} we study the evolution of the effective gravitational coupling, which is then used in Section~\ref{Sec:lineargrowth} to investigate the linear growth of structures through the evolution of the growth factor and the product of the growth rate and root mean square of matter fluctuations. We also provide theoretical predictions for the sign of the Integrated Sachs-Wolfe (ISW)-galaxy cross correlation in Section~\ref{Sec:ISWgal}. Finally, we conclude in Section~\ref{Sec:conclusion}.

\section{\boldmath $f(Q)$ gravity}\label{Sec:fQbasics}

General Relativity is a metric theory of gravity and as such the connection is metric-compatible and symmetric. However, two alternative approaches can be considered to characterize the space-time: non-metricity and torsion, giving up the first and second assumptions, respectively. It has been shown~\cite{BeltranJimenez:2019tjy} that in flat space-time the Einstein-Hilbert action,
the teleparallel ($\int{d^4x\sqrt{-g}\,T}$ \cite{Aldrovandi:2013wha}) and symmetric teleparallel ($\int{d^4x\sqrt{-g}\,Q}$ \cite{BeltranJimenez:2017tkd}) actions are three different representations of the same underlying theory. Nevertheless, MG theories based on non-linear extensions of both the non-metricity scalar, $Q$, namely $f(Q)$ gravity, as well as torsion ($T$), namely $f(T)$ gravity \cite{Bahamonde:2021gfp}, can be considered. These, unlike the previous case, are not equivalent. In this work we will focus on the $f(Q)$ formulation given the absence of strong coupling problems for the FLRW background~\cite{BeltranJimenez:2019tme} compared to $f(T)$ gravity and its increasing interest in the cosmological framework \cite{Barros:2020bgg,Anagnostopoulos:2021ydo,Atayde:2021ujc}, as discussed in the Introduction. Interestingly, $f(Q)$ gravity introduces at least  two additional scalar propagating degrees of freedom which disappear around maximally symmetric backgrounds~\cite{BeltranJimenez:2019tme}.

Let us introduce the action  of  $f(Q)$ gravity which reads\footnote{Comparing our action (\ref{eq:action}) with the one in Ref.~\cite{BeltranJimenez:2019tme}, we have performed the following replacement $f(Q)\rightarrow \frac{1}{\kappa^2}\left(Q+f(Q)\right)$, because this form better fits our purpose.} ~\cite{BeltranJimenez:2019tme}
\begin{equation} \label{eq:action}
S=\int d^4x\sqrt{-g}\left\{-\frac{1}{2\kappa^2}\left[Q+f(Q)\right]+L_m(g_{\mu\nu},\chi_i)\right\},
\end{equation}
where $\kappa^2 = 8 \pi G_N$, $G_N$ is the Newtonian constant, $g$ is the determinant of the metric $g_{\mu\nu}$ and $Q$ is the non-metricity scalar defined as
\begin{equation}
Q=-Q_{\alpha\mu\nu}P^{\alpha\mu\nu}\,,
\end{equation} 
where the non-metricity tensor is:  
\begin{equation}
Q_{\alpha\mu\nu}=\nabla_\alpha g_{\mu\nu}\,,
\end{equation}
and 
\begin{equation}
P^{\alpha}_{\phantom{\alpha}\mu\nu}=-L^{\alpha}_{\phantom{\alpha}\mu\nu}/2+\left(Q^\alpha-\tilde{Q}^\alpha\right)g_{\mu\nu}/4-\delta^\alpha_{(\mu}Q_{\nu)}/4\,,
\end{equation}
with $Q_\alpha=g^{\mu\nu}Q_{\alpha\mu\nu}$, $\tilde{Q}_\alpha=g^{\mu\nu}Q_{\mu\alpha\nu}$ and $L^\alpha_{\phantom{\alpha}\mu\nu}=(Q^\alpha_{\phantom{\alpha}\mu\nu}-Q_{(\mu\nu)}^{\phantom{(\mu\nu)} \alpha})/2$. Finally, $f(Q)$ is a general function of the non-metricity scalar and as usual $L_m$ is the matter Lagrangian for all matter fields, $\chi_{i}$.  

In flat space the action (\ref{eq:action}) is equivalent to GR for $f=0$~\cite{BeltranJimenez:2019tjy} since the symmetric teleparallel action is recovered. Thus any modification to GR can be seen for $f\neq0$. From the above action it is also possible to compute the equation for tensor perturbations which is characterized by a shift in the friction term due to the time derivative of an effective Planck mass, given by $M^2_{\rm eff}=1+f_Q$ where $f_Q\equiv df/dQ$~\cite{Jimenez:2019ovq}. From this follows that in order to guarantee the absence of a ghost instability one has to impose $1+f_Q>0$. We will assume this condition in our investigation.

\section{Designer approach} \label{Sec:designer}

Let us assume a flat, homogeneous and isotropic Universe described on the background by the FLRW metric:
\begin{equation}
    ds^2=-dt^2+a(t)^2\delta_{ij}dx^idx^j\,,
\end{equation} 
where $a(t)$ is the scale factor. On this background the non-metricity scalar simply reduces to $Q = 6H^2$ \cite{BeltranJimenez:2017tkd,Jimenez:2019ovq} where $H\equiv\dot{a}/a$ is the Hubble parameter and the dot stands for a derivative with respect to cosmic time, $t$. 

The equations of motion on this background are:
\begin{eqnarray}
    &&H^2 + 2 H^2 f_Q - \frac{1}{6} f = \frac{\kappa^2}{3} \rho_{\rm m} , \label{eq:FriedEq} \\
  &&(12H^2f_{QQ}+f_Q+1)\dot{H} = -\frac{\kappa^2}{2}(\rho_{\rm m}+p_{\rm m})\,,
\end{eqnarray}
with $\rho_{\rm m}=\Sigma_i\rho_i$ and $p_{\rm m}=\Sigma_ip_i$ being respectively the sum of the energy density, $\rho_i$, and pressure, $p_i$, of the matter components, which satisfy the continuity equation for perfect fluids:
\begin{equation}
    \dot{\rho}_i+3H(\rho_i+p_i)=0\,.
\end{equation}
Hereafter we will consider the relation $p_i=w_i\rho_i$, with $w_{c,b}=0$ for baryons (b) and cold dark matter (c), and  $w_r=1/3$ for radiation (r). For the times of interest ($a \in [10^{-2},1]$) the matter components behave as non-relativistic matter fluids.

In order to solve the above modified Friedmann equation we have to fix either the functional form of $f(Q)$ or that of $H$.  A common practice is to select the form of $f(Q)$  and then solve the system to find the  corresponding expansion history, $H$. In this work we will follow the opposite approach: we will fix $H$ and solve the system to find the corresponding functional form for $f(Q)$. Hence we will apply the so-called \textit{designer} approach \cite{Song:0610532,Pogosian:0709}. 

We define the dimensionless variables
\begin{equation}
    E \equiv \frac{H^2}{H_0^2}, \ \ \ \ y \equiv \frac{f(Q)}{H_0^2},  
\end{equation}
where $H_0$ is the present day value of the Hubble parameter, and we fix the expansion history such that
\begin{equation}
    E = E_r + E_c + E_b + E_{Q}\, .
\end{equation}
In the latter, $E_i = \rho_i / \rho_{crit}$ with $\rho_{crit} \equiv 3H_0^2 / \kappa^2 $ and the energy density parameter for the effective dark energy component associated to $Q$, $E_Q$, is given by
\begin{equation}\label{eq:EnDenQ}
    E_{Q} = ( 1 - \Omega_{\rm m,0} ) \ \mbox{exp} \left[ - 3 \ln a + 3 \int_{a}^{1} w_{Q} (\tilde{a}) d \ln \tilde{a} \right] ,
\end{equation}
where $\Omega_{\rm m,0}$  is the present time value of the density parameter $\Omega_{\rm m}\equiv \Sigma_i \rho_i/(3H_0^2 / \kappa^2) $  and $w_Q (a)$ is the equation of state parameter for an effective dark energy component.

Now we can re-work the Friedmann equation (\ref{eq:FriedEq}) as a first-order differential equation
\begin{eqnarray}\label{eq:diffeq}
  y' - \frac{Q'}{12 E H_0^2} y = - \frac{Q'}{2 E H_0^2} E_{Q} \,,
\end{eqnarray}
where the prime denotes a derivative with respect to $\ln a$. In order to solve this equation we set  initial conditions (ICs) at early time, $a_i$, when we consider the contribution of the effective dark energy component negligible compared to the matter one. The resulting equation becomes  homogeneous and is satisfied by the ansatz $y \propto \exp(p \ln a)$, where $p$ is obtained by solving the homogeneous equation yielding
\begin{equation}
    p = - \frac{3+4r_i}{2 (1+r_i)} ,
\end{equation}
with $r_i = a_{\rm eq}/a_i$ and $a_{\rm eq}$ being the value of the scale factor at matter-radiation equality.

Then we look for the particular solution, $y_p$, when $E_Q\neq0$, which is obtained by substituting $y_p = A_p E_Q$ in eq.~(\ref{eq:diffeq}). The amplitude of the particular solution is then 
\begin{equation}
    A_p = \frac{6 p}{3(1+w_Q) + p}\,,
\end{equation}
leading to the following initial condition for $y$:
\begin{equation} \label{eq:initcond}
   y_i = A  a_i^p + \frac{6 p}{3(1+w_Q) + p} E_Q  \ ,
\end{equation}
where $A$ is an arbitrary constant. We note that  the amplitude of the ICs is not an appropriate parameter to define the family of solutions of eq.~(\ref{eq:diffeq}).  Indeed a change in the initial time, $a_i$, would correspond to a rescaling in $A$ in order to  obtain the same solution. This would make it difficult to obtain any theoretical prediction based on $A$, given that to different $a_i$ would correspond different values of $A$ with the same behaviour. As such, in order to avoid this contingency, we define another quantity which will characterize the family of solutions:
\begin{equation}
\alpha=\frac{A \,a_i^p}{\sqrt{6E(\ln a_i)}}\,,
\end{equation}
which follows from the fact that $y/\sqrt{E}$ is constant at early time (see eq.~(\ref{eq:diffeq})). In this case, $\alpha$ does not have to change when $a_i$ is altered given that it can be compensated by a rescaled value of $A$.

With this set up, we are left with few more parameters to consider: the constant $\alpha$ and the parameters entering in the equation of state, $w_Q (a)$, which basically fix the background expansion. In this work we will explore three options for dealing with $w_Q$:
\begin{enumerate}
    \item \textbf{\boldmath Constant $w_Q$}: $w_Q = -1$
            
        The simplest case to be considered is that of having $w_Q = -1$ which would reproduce an exact $\Lambda$CDM background expansion history. In this situation, the energy density for $Q$ $ \left( E_Q = 1 - \Omega_{ \rm m,0} \right)$ becomes a constant  and the differential eq.~(\ref{eq:diffeq}) can be analytically solved. We find:
        \begin{equation} \label{eq:ansol}
        y  = \frac{y_i - 6 E_Q}{\sqrt{E(\ln a_i)}} \sqrt{E} + 6 E_Q \, ,            \end{equation}
        or in terms of $f(Q)$:
        \begin{equation} \label{eq:fQan}
        f(Q) =  \alpha H_0\sqrt{Q} + 6H_0^2 E_Q \, ,
        \end{equation}
        which is consistent with what was found in Ref. \cite{Jimenez:2019ovq}, upon the following identification: $M=\alpha H_0$ and $C = 6H_0^2 E_Q$. Let us note that combined with the choice $\alpha=0$, this case corresponds exactly to a $\Lambda$CDM behaviour at perturbation level as well. 

    \item  \textbf{\boldmath Constant $w_Q$}: $w_Q=w_0$

        The second case we consider has a constant $w_Q=w_0$ with $w_0\neq -1$. In this situation, $E_Q$ becomes
        \begin{equation}
        E_Q (a) = (1 - \Omega_{\rm m,0}) \, a^{  -3 \left( 1 + w_0 \right) },
        \end{equation}
        which means it is no longer a constant as long as $w_0 \neq -1$. With this choice of the background it is not  possible to find an analytical solution for $f(Q)$. We will use this case to understand the impact of changing the equation of state parameter on some observational features. For the purpose of visualizing these features, in the following we will select four values for $w_0$, namely $\{-1.15, -1.05, -0.95, -0.85\}$.
    \item \textbf{\boldmath Time-dependent $w_Q$}: $w_Q(a)$ 
    
        The third case we consider assumes a fast varying equation of state \cite{DeFelice:1203}. In this case, $w_Q$ evolves as
        \begin{equation}
            w_Q (a) = w_p + (w_0 - w_p) \, \frac{a \left[ 1 - \left( a/a_t \right)^{1/\tau} \right]}{1-a_t^{-1/\tau}} ,
        \end{equation}
        where $w_p$ and $w_0$ are the values of $w_Q$ in the asymptotic past and at present time, respectively, and $a_t$ and $\tau$ are the time and the width of the transition. Additionally, by definition we have $a_t > 0$ and $\tau > 0$.
        In this case, the energy  density of the effective dark energy component becomes:
        \begin{equation}
           E_Q = \left( 1 - \Omega_{\rm m,0} \right) a^{  -3 \left( 1 + w_p \right)}  e^{g(a)}  ,
        \end{equation}
        with
        \begin{eqnarray}
        g(a) &=&  \frac{3(w_0 - w_p)}{(1 - {a_t}^{-1/\tau})(\tau + 1)}  \left[ \left( 1 - {a_t}^{-1/\tau} \right) \tau \right.\nonumber\\
        &+&\left.1+ a  \left[ \left( \left( a / a_t \right)^{1/\tau} - 1 \right] \tau - 1 \right)  \right] .
        \end{eqnarray}
        As such, we will have four free parameters: $\{ w_0, w_p, \\a_t, \tau \}$. In order to reduce the possible combinations, and because we are already exploring cases where $w_Q=w_0$, we shall fix $w_0 = -1$. Additionally, we  want to keep the transition time always prior to present day, meaning keeping $a_t < 1$. We will then use the following sets of values for the remaining parameters: $\{ w_p, a_t, \tau \} ={\{ -1.15, 0.5, 10 \}}$ and $\{ w_p, a_t, \tau \} ={\{ -0.95, 0.5, 10 \}}$, we will refer to them as M1 and M2 respectively. We do not consider cases in which $a_t$ and $\tau$ vary because we have verified they do not have any sizable impact on the phenomenology we are interested in. Let us note that with these choices the associated $E_Q(\ln a_i)$ remains subdominant. \end{enumerate}

In the present analysis we will use the following cosmological parameters \cite{Planck:2018vyg}: $H_0= 67.32$ km s$^{-1}$Mpc$^{-1}$, ${\Omega_{c,0}= 0.265}$, $\Omega_{b,0}= 0.049 $, $\Omega_{r,0}= 3.769 \times 10^{-5}$, and for the initial time we select $a_i\sim 10^{-2}$.

Let us also stress that some of the cases we consider have $w_Q<-1$. While this behaviour of the equation of state usually generates ghost instabilities, for the $f(Q)$ theory analysed in this work, the ghost instability is avoided by choosing $f_Q>0$ at any time. This condition inevitably reflects on the parameter space of the free parameters defining the $w_Q$ equation of state.  Therefore, in performing our analysis we have verified that the set of parameters we chose satisfy the no-ghost requirement.

\section{The effective gravitational coupling} \label{Sec:effectivecoupling}

The next step will be to investigate the evolution of the scalar perturbations at linear scales. Firstly, we will review the main equations which involve the use of the linear perturbation theory, then we will show how the gravitational interaction is modified with respect to the standard $\Lambda$CDM scenario and its dependence on the background assumptions.

Let us consider the perturbed line element in Newtonian gauge:
\begin{equation}
    ds^2=-(1+2\Psi)dt^2+a^2(1-2\Phi)\delta_{ij}dx^idx^j\,,
\end{equation}
where $\Phi(t,x_i)$ and $\Psi(t,x_i)$ are the two gravitational potentials.
For MG theories a model-independent framework is usually adopted to relate the gravitational potentials to the linear matter density perturbations $\delta \rho_{\rm m}$ ~\cite{Amendola:2007rr,PhysRevD.81.083534,Silvestri:2013ne,2010PhRvD..81j4023P,Amendola:2019laa}. This framework encodes the deviations from GR into two phenomenological functions, namely the {\it effective gravitational coupling}, $\mu$,  and the {\it light deflection parameter}, $\Sigma$, which enter in the Poisson and lensing equations, respectively. In Fourier space the latter read:
\begin{eqnarray}
\label{eq:muSigma}
&&-\frac{k^2}{a^2} \Psi = 4 \pi G_N \,\mu(a, k)  \rho_\mathrm{m} \delta_{\rm m}\,, \\
&&-\frac{k^2}{a^2} (\Psi+\Phi) = 8\pi G_N\Sigma(a,k) \rho_\mathrm{m}\delta_{\rm m}\,, \label{eq:lenseq}
\end{eqnarray}
where $\delta_{\rm m}=\delta \rho_{\rm m}/\rho_{\rm m}$ is the density contrast and $k$ is the wavenumber. Therefore, $\mu$ encodes the deviations of the gravitational interaction on the clustering of matter with respect to $\Lambda$CDM, while $\Sigma$ measures the deviation in the lensing gravitational potential, $\phi_{len}=(\Phi+\Psi)/2$. The $\Lambda$CDM model is recovered when $\mu=\Sigma=1$.

In order to map the $f(Q)$ gravity within the above formalism one can employ the quasi-static approximation, which is a valid assumption for perturbations deep inside the Hubble radius. Following this, one can find that the gravitational potentials are equal as in GR (i.e. $\Phi=\Psi$) and that the two above equations match~\cite{Jimenez:2019ovq}:
\begin{equation}
    -k^2\Psi= \frac{4 \pi G_N}{1+f_Q}a^2\rho_{\rm m} \delta_{\rm m}\,.\label{eq:Poisson}
\end{equation}
The effective gravitational coupling is then defined as:
\begin{equation}\label{eq:mu}
     \mu(a)=\frac{1}{1+f_Q}\,.
\end{equation}
When $f_Q\rightarrow 0$ the $\Lambda$CDM behaviour is recovered, i.e. $\mu=1$. 
From eq.~(\ref{eq:mu}) immediately follows that since $M^2_{\rm eff}=1+f_Q>0$ due to stability requirements, then $\mu$ is always positive. Additionally, the cases $\mu<1$ and $\mu>1$ correspond respectively to a weaker and stronger gravitational interaction compared to $\Lambda$CDM. Furthermore, the light deflection parameter is then equal to the effective gravitational coupling and we notice that the scale dependence, i.e., the dependence on $k$, disappears.

\begin{figure}[t!]
\centering
\includegraphics[scale=0.56]{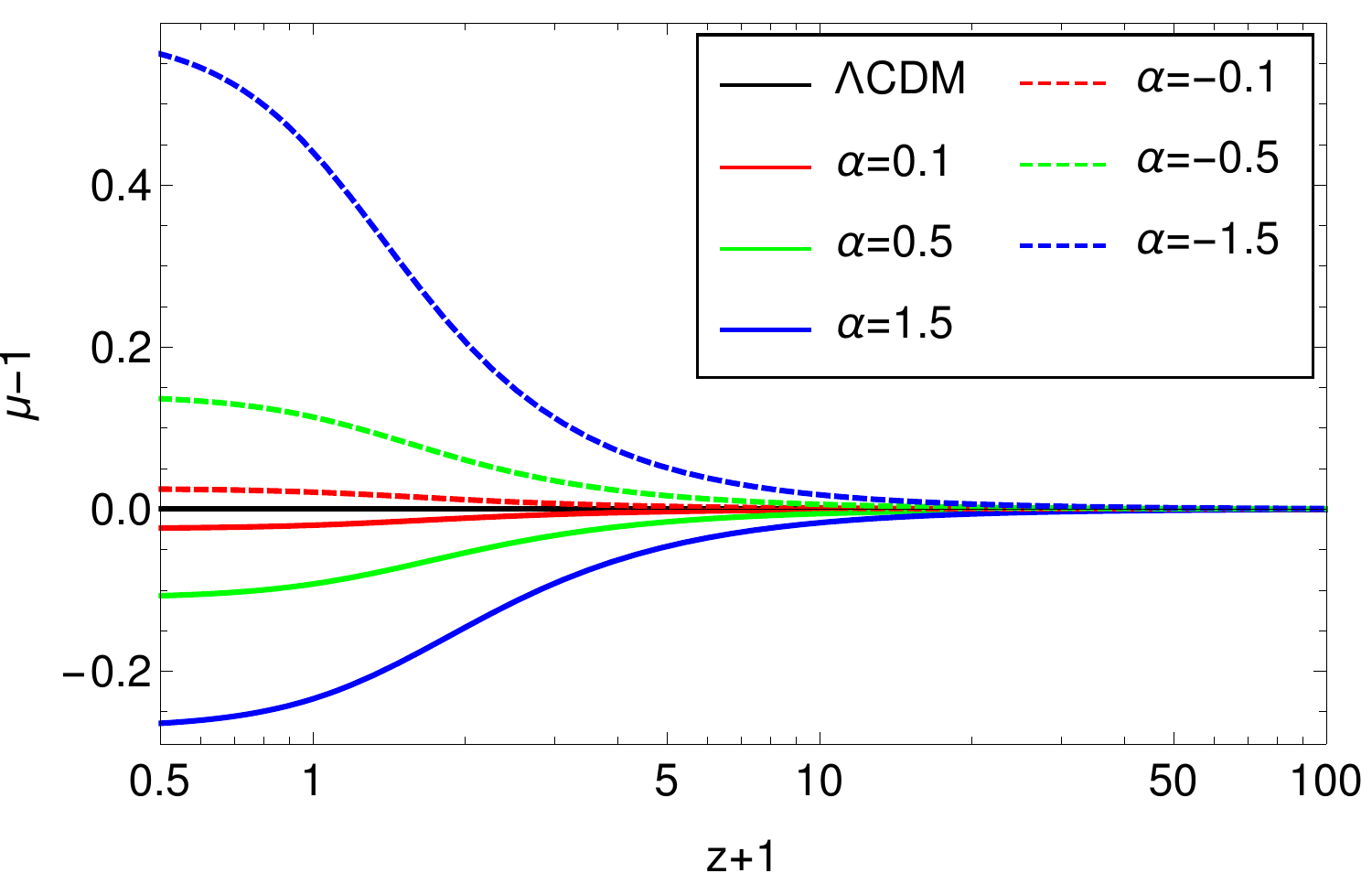}
\caption{Evolution of $\mu - 1$ as a function of redshift $z$ for models with an  exact $\Lambda$CDM background and different values of $\alpha$. \label{fig:mupltsw-1}}
\end{figure}

From eq.~(\ref{eq:mu}) it becomes clear that depending on the functional form of $f(Q)$, the evolution of the linear perturbations will change accordingly. In our investigation the form of $f(Q)$ will be determined by the use of the designer approach. Given this, in the following we will study the evolution of $\mu - 1$, i.e. of the difference between the effective gravitational couplings of $f(Q)$ and $\Lambda$CDM, for the different background expansion histories discussed in Section~\ref{Sec:designer}. This investigation is of particular interest because a modified effective gravitational coupling  will impact the growth and distribution of structures in time and space. Additionally, since $\mu = \Sigma$, we are also studying the modifications of the Weyl potential, $\phi_{len}$, by the light deflection parameter, which will in turn change the lensing of light and modify the ISW effect. The latter is indeed sourced by the time derivative of the  Weyl potential. As such, in $f(Q)$ all these effects are encoded in $\mu$.

The case with $w_Q=-1$ has already been presented and discussed in Ref. \cite{Frusciante:2021sio}.
Briefly, we recall the main features. From Figure~\ref{fig:mupltsw-1}, we see that sizable deviations from GR appear at $z<10$ and then grow towards present time. Values of $\alpha>0$ result in a weaker gravity compared to GR ($\mu - 1<0$) while the opposite holds for $\alpha<0$. Additionally, larger values of $|\alpha|$ lead to larger deviations from the $\Lambda$CDM behaviour, however for the same $|\alpha|$ but opposite sign, we notice that the modification corresponding to the $\alpha<0$ gives rise to a larger deviation compared to its positive counterpart. 

\begin{figure}[t]
  \centering
  \includegraphics[scale=0.56]{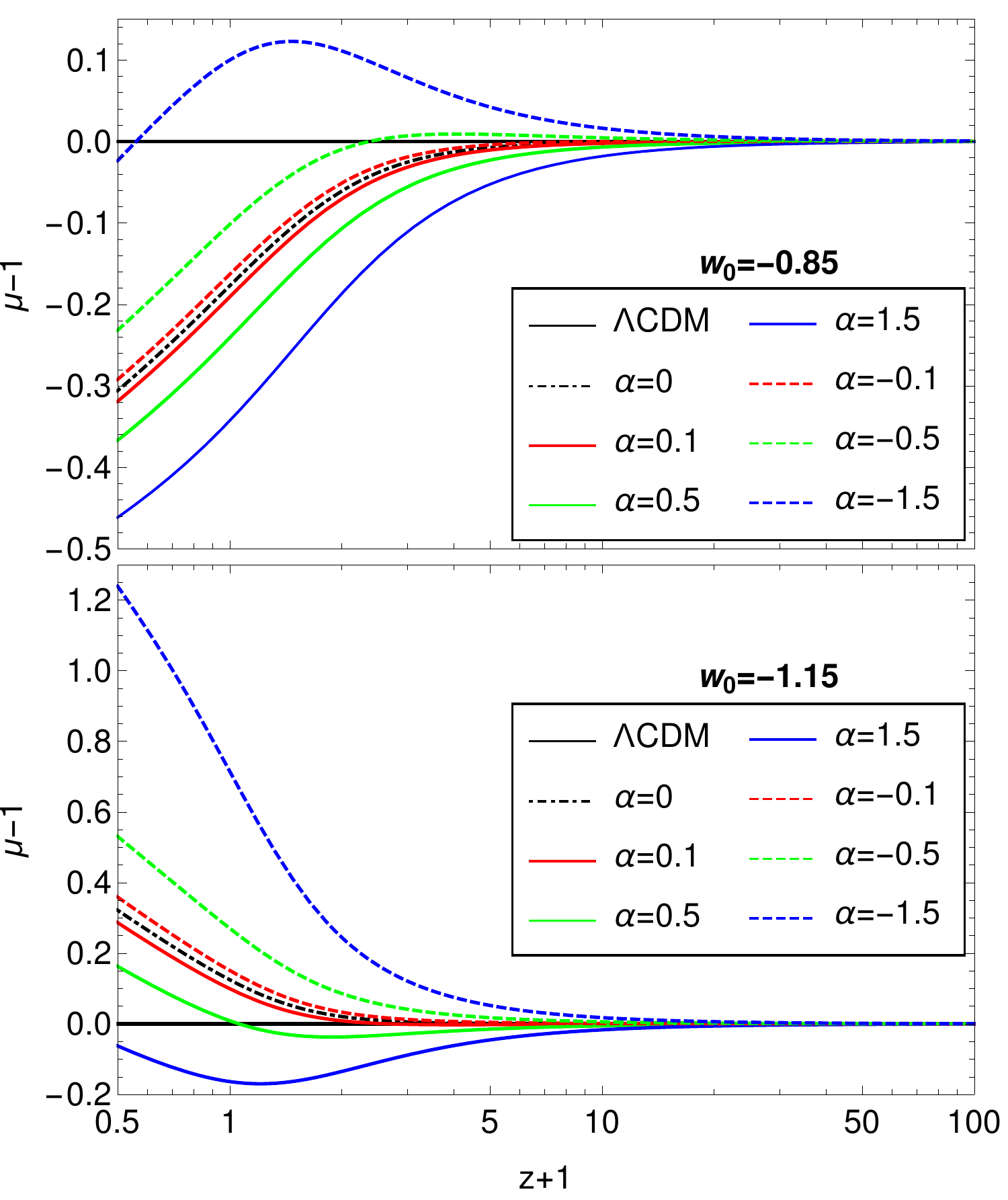}
\caption{Evolution of $\mu - 1$ as a function of redshift $z$ for different $\alpha$ values with fixed $w_0 = -0.85$ (top panel) and $w_0=-1.15$ (bottom panel). }
\label{fig:mupltsVwvwQ}
\end{figure}

\begin{figure}[t!]
  \centering
  \includegraphics[scale=0.56]{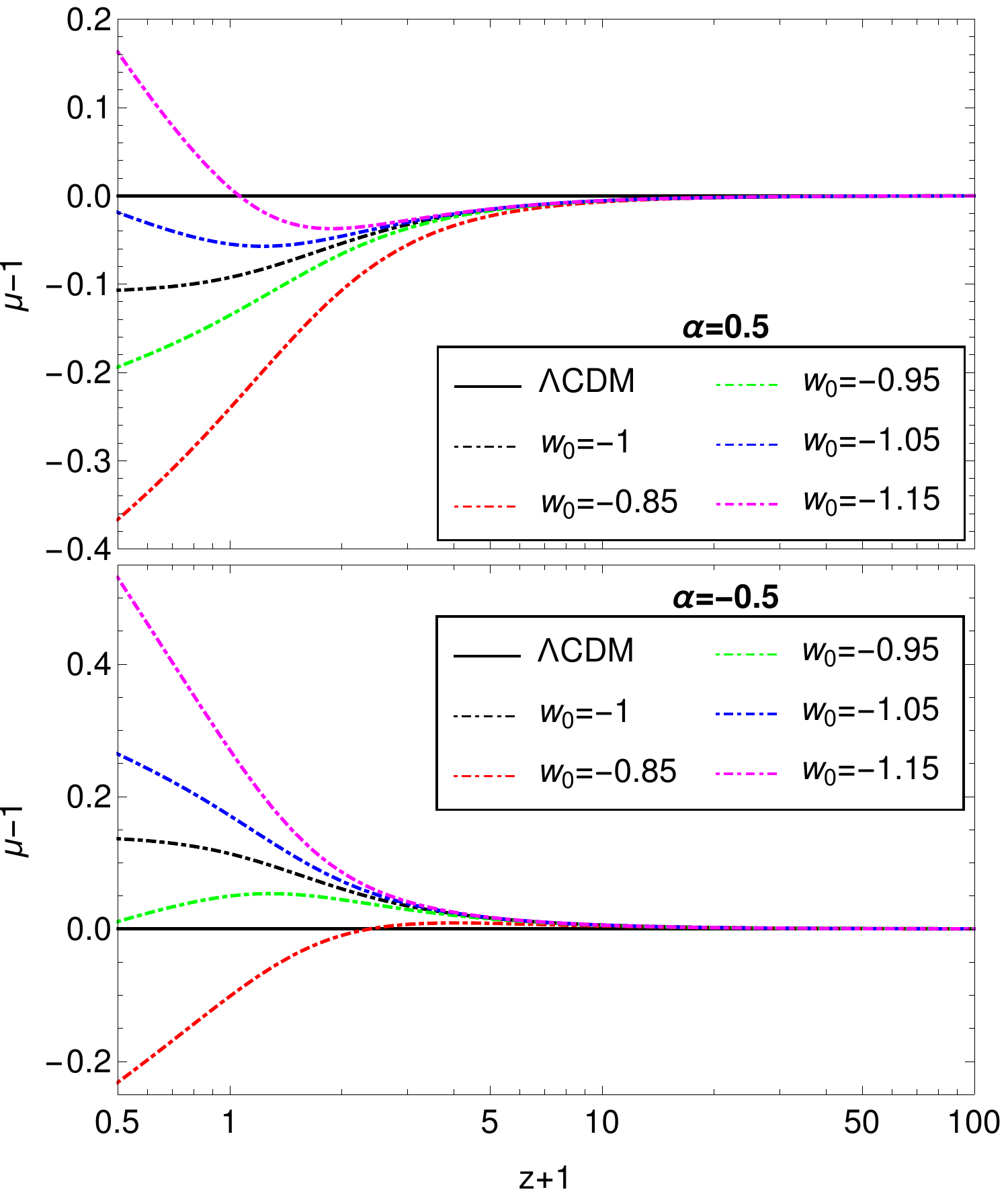}
\caption{Evolution of $\mu - 1$ as a function of redshift $z$ for different values of $w_0$ with fixed  $\alpha=0.5$ (top panel) and $\alpha = -0.5$ (bottom panel).}
\label{fig:mupltsVAvwQ}
\end{figure}

Let us now consider the case with  $w_Q=w_0$. In Figure~\ref{fig:mupltsVwvwQ} we show the results when $w_0=-0.85$ (top panel) and $w_0=-1.15$ (bottom panel) for some given values of $\alpha$. When $w_0=-0.85$, regardless of the sign of $\alpha$, the $\mu-1$ behaviour goes toward negative values at small $z$, i.e. toward a weaker gravitational interaction. In order to have a stronger gravitational interaction at present time the value of $|\alpha|$ has to be very large and of negative sign ($\alpha=-1.5$). The opposite holds if we consider the phantom values of $w_0$ (bottom panel). In this case, the gravitational interaction is stronger toward present time and the weaker gravity is realized when  the value of $|\alpha|$ is very large and of positive sign ($\alpha=1.5$). According to these features, for a given value of $\alpha$ the direction of the modifications (toward weaker/stronger gravity) is dictated by the value of $w_0$, while $\alpha$ mostly impacts on the amplitude of the deviation with respect to $\mu=1$. This is more clear in Figure~\ref{fig:mupltsVAvwQ}, where we fix $\alpha$ and vary $w_0$. 

\begin{figure}[t!]
\centering
  \includegraphics[scale=0.56]{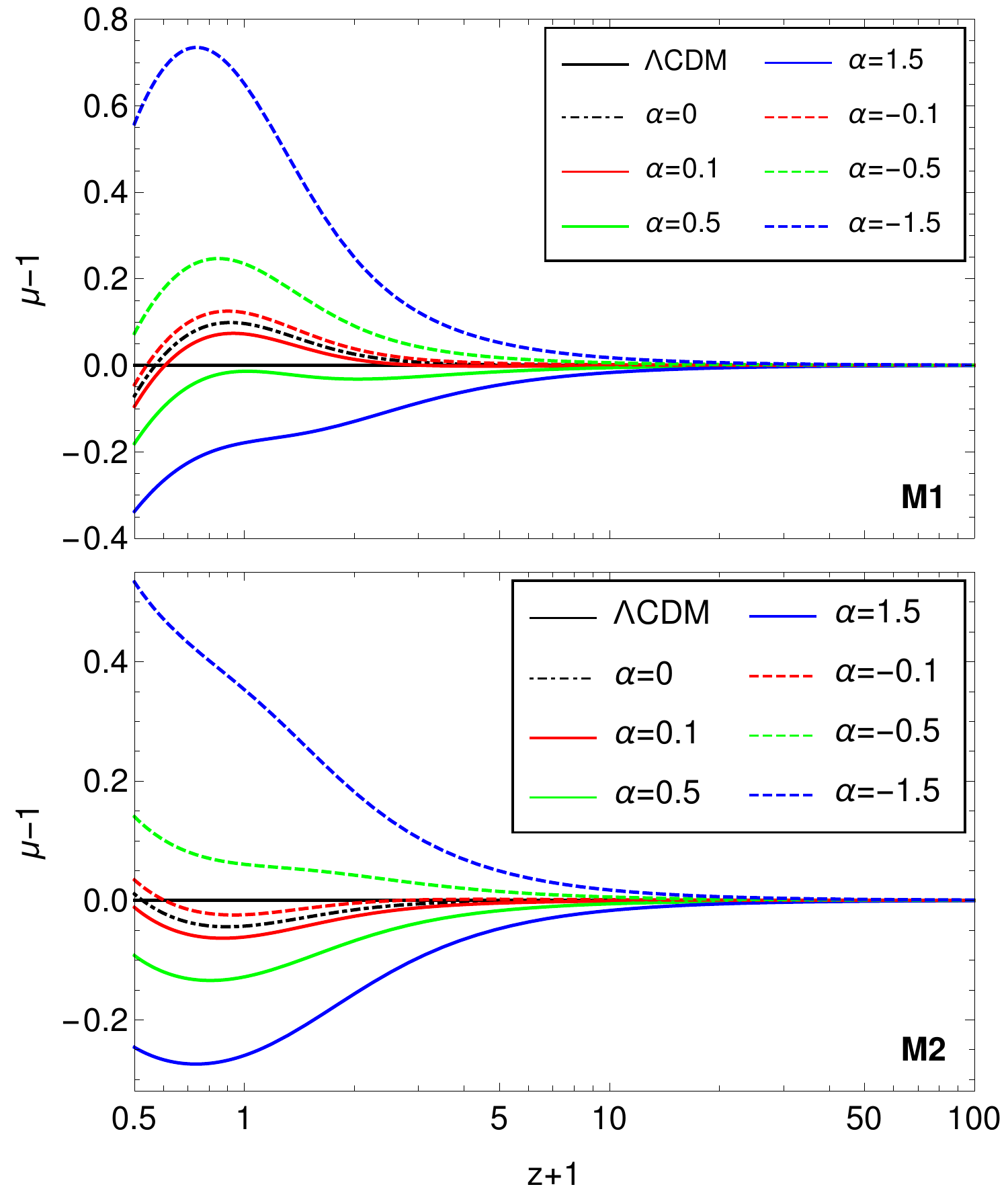}
\caption{Evolution of $\mu - 1$ as a function of redshift $z$ for the time-dependent $w_Q$ (M1 top panel and M2 bottom panel).}
\label{fig:muplts1DwQ}
\end{figure}

Finally, we discuss the behaviour of $\mu -1$ for the time-dependent $w_Q$ model, for which we show the result in Fig. \ref{fig:muplts1DwQ}.  As noted for the case with an exact $\Lambda$CDM background, most of the models with a negative $\alpha$ parameter have $\mu>1$ whereas the opposite holds for positive $\alpha$. However, while in the former case there is a net distinction, when $w_Q$ has fast transitions the models corresponding to small values of $|\alpha|$ can have stronger or weaker gravity depending on the value of $w_p$ at early time. In particular, if $w_p>-1$ such as in the M2 case (bottom panel), then $\mu<1$. Alternatively, if $w_p<-1$ we then have $\mu>1$ as shown in the top panel.  Finally, if we look at the behaviours of $\mu-1$ in the future, we notice that all models that have $w_p < -1$ (top panel) will have $\mu - 1$ going toward negative values, while when $w_p > -1$ it is the opposite. This is due to the fact that after present time the M1 model undergoes another matter dominated era, while M2 stays in the dark fluid dominated one. 

In summary, the behaviour of the effective gravitational coupling  depends strongly on the assumed background expansion history.  However, let us notice that while the background expansion history only depends on the parameters defining $w_Q$, the effective gravitational coupling depends on both $w_Q$ and $\alpha$, showing a degeneracy between the parameters in some cases. In particular, we find that a preference for a weaker/stronger gravitational interaction can be secured by a proper choice of either the present time value of $w_Q$ or its asymptotic past one. In detail, phantom behaviours at late time  and early time modifications with $w_p>-1$ prefer a stronger gravity, the  weaker gravitational interaction is instead present in the opposite situations. The  role of $\alpha$ is to define the amplitude of the deviation with respect to $\Lambda$CDM, its negative and positive values contribute to enhance or suppress the effective gravitational coupling with respect to the standard scenario. Extremely large values of $|\alpha|$ can even reverse the strength of the gravitational interaction. These peculiar patterns have some immediate consequences on the clustering of matter and on how light travels over cosmological distances. We expect indeed that a stronger gravitational interaction will lead to  an enhanced matter and  lensing potential auto-correlation power spectra. As an example, we refer the reader to Ref. \cite{Frusciante:2021sio}, where an investigation of signatures of $f(Q)$ gravity, when the background is assumed to be $\Lambda$CDM, has been performed by looking at the matter power spectrum, the lensing effect on the CMB angular power spectrum, CMB temperature anisotropies and GWs luminosity distance compared to the standard electromagnetic one.

We note that the equation of state with time-dependent $w_Q$ can actually cross $w_Q=-1$. When the cross happens (for values of the parameters for which the condition $f_Q>0$ is satisfied) we found that the general behaviour of $\mu(z)-1$ as shown in Fig.~\ref{fig:muplts1DwQ} is still the same but with a difference in the amplitude. The latter, depending on the value of $w_0$, can be larger or smaller than the case with $w_0=-1$ we have considered, following our findings illustrated in Fig.~\ref{fig:mupltsVAvwQ}. We did not show this case because a change in $w_0$ for the time-dependent equation of state follows the results for the constant case, making some of the analysis redundant. We will not consider this case in the following sections, but all the other phenomenological quantities we compute will be modified accordingly, following the same pattern.

In order to get a glimpse in the phenomenology of the designer $f(Q)$ gravity related to the effective gravitational coupling, we dedicate the following section to the investigation of the linear growth of structures. Furthermore, the latter combined with the late-time ISW effect largely impact the sign of the cross-correlation function between the galaxy density and the temperature of the CMB. As such, its study can be informative regarding the parameter space to explore in order to exclude a negative ISW-galaxy cross-correlation which is observationally disfavored.

Before concluding this section, we notice that an accurate combination of data may help  breaking the existing degeneracy between $w_Q$ and $\alpha$. If one is able to constrain with high accuracy the equation of state, using for example SNIa and BAO data, then GC, RSD as well as CMB data can be used to probe the effects of $\mu$ on some cosmological observables at perturbation level and as such set accurate bounds on $\alpha$ as well. An independent estimation of $\alpha$ may also come from measurements of standard sirens at future GWs detectors \cite{Frusciante:2021sio}.  

\section{Linear growth of structures} \label{Sec:lineargrowth}

In this section we will investigate how the effective gravitational coupling impacts the linear growth of structures. In particular we will study the product of the growth rate and the root mean square of matter fluctuations, $\tilde{f}\sigma_8$, as it has a direct connection with data and hence offers a possibility to explore the parameter space given by $\{w_Q,\alpha\}$.

The growth of linear matter perturbations is modified in the $f(Q)$ theory due to a modified Poisson eq.~(\ref{eq:Poisson}) as follows:
\begin{equation} \label{eq:lingrtheq}
    \delta_{\rm m}'' + \left( 2 + \frac{H'}{H} \right) \delta_{\rm m}' - 4\pi G_N \mu (a) \, \rho_{\rm m} \delta_{\rm m} = 0 .
\end{equation}
We now solve the above equation with ICs set as follows:  $\delta_{\rm m} (a_i) = a_i$ and $\delta_{\rm m}' = a_i$.  Then, we compute the linear growth factor $D(a) = \delta_{\rm m }(a) / \delta_{\rm m} (a=1)$ for the three scenarios analyzed in this work.

The linear growth factor for the case with an exact $\Lambda$CDM background expansion history has been investigated in Refs. \cite{Barros:2020bgg,Frusciante:2021sio}, where it has been found that models with $\alpha > 0$ and $\mu - 1 <0$ show an enhanced growth factor with respect to the $\Lambda$CDM scenario, whereas for $\alpha < 0$ and $\mu - 1 >0$, $D(a)$ is suppressed. 

\begin{figure}[t!]
  \centering
  \includegraphics[scale=0.56]{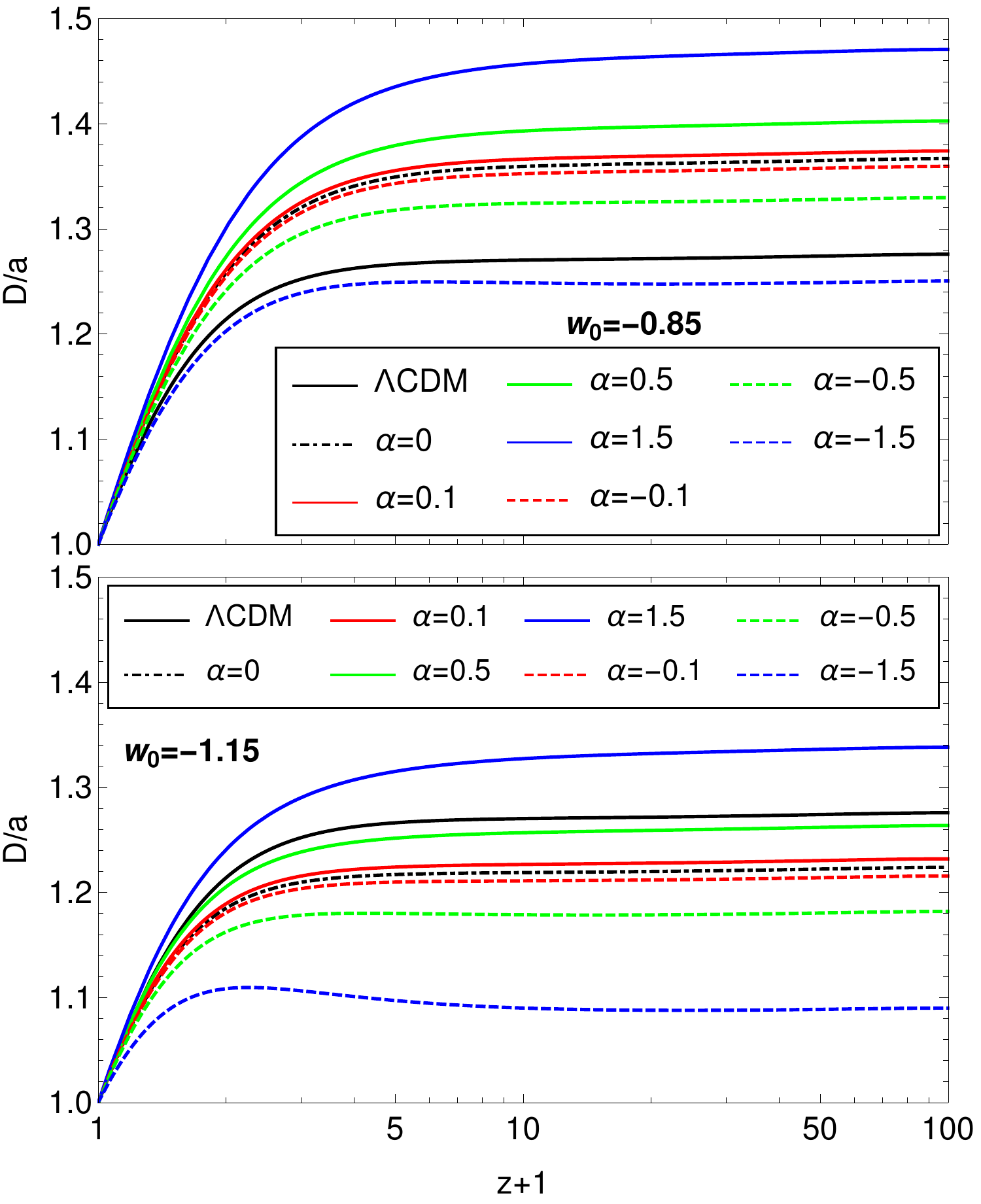}

\caption{Evolution of the linear growth factor $D$ normalized to unity today and divided by the scale factor $a$ as a function of redshift $z$ for models with different $\alpha$ and fixed $w_0=-0.85$ (top panel) and $w_0=-1.15$ (bottom panel). The $\Lambda$CDM model (solid, black line) is also included for comparison. }
\label{fig:DpltsVWvwQ}
\end{figure}

When changing the background expansion to have $w_Q=w_0$, the growth factor is shifted in amplitude as shown in Figure~\ref{fig:DpltsVWvwQ}. Specifically, when $w_0>-1$ (top panel) the growth factor is enhanced (black dot-dashed line) with respect to $\Lambda$CDM. Then, depending on the values of $\alpha$, $D$ can be enhanced or suppressed compared to $\Lambda$CDM. Large negative values of $\alpha$ suppress the linear growth factor. On the contrary, when the equation of state has a phantom behaviour (bottom panel), the growth factor is mostly suppressed, the large positive $\alpha$ being the exception. That is due to the fact that in the former case the models mostly show a weaker gravity compared to the case where $w_0<-1$. The two exceptions are indeed the models with the largest negative and positive values of $\alpha$ respectively. This is consistent with the $\mu$ behaviour analysed in the previous section.

\begin{figure}[t!]
  \centering
  \includegraphics[scale=0.56]{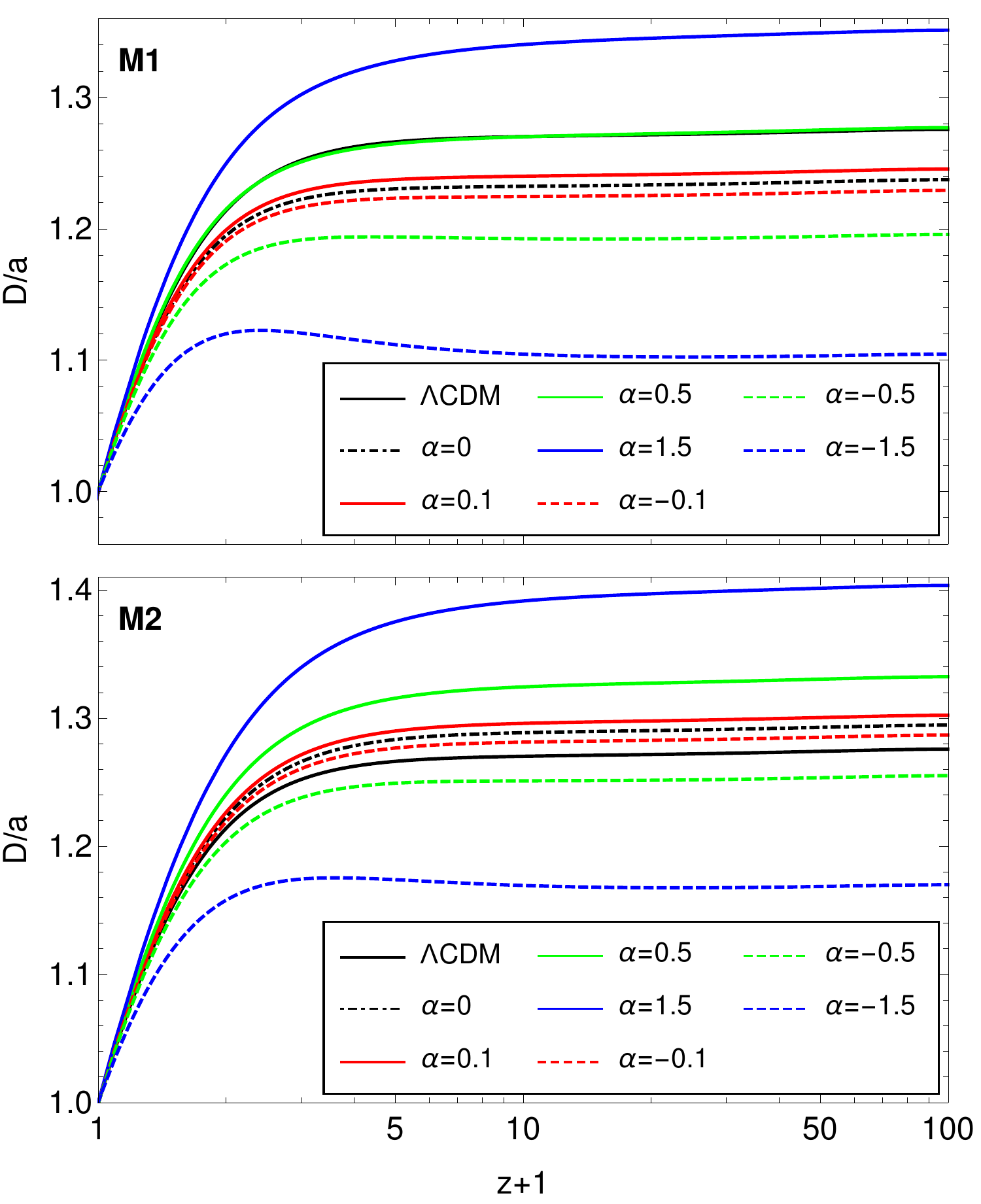}
\caption{Evolution of the linear growth factor $D$ normalized to unity today and divided by the scale factor $a$ as a function of redshift $z$ for the time-dependent equation of state $w_Q$ : M1 (top panel) and M2 (bottom panel). The $\Lambda$CDM (solid, black line) is also included for comparison.}
\label{fig:DpltsDwQ}
\end{figure}

For the time-dependent $w_Q$ case, the behaviour of the growth factor is shown in Figure~\ref{fig:DpltsDwQ} for M1 (top panel) and M2 (bottom panel) and different values of $\alpha$.  We can notice that a different evolution in the early time can change the amplitude of the growth factor. Indeed having $w_Q=-1$ would make a net separation between $\alpha$ positive and negative which would correspond respectively to an enhancement/suppression of the growth with respect to $\Lambda$CDM, while here a dynamical evolution at early time breaks this pattern by changing the amplitude of the growth factor. In particular, a phantom $w_p$ (M1) damps the amplitude of the growth factor and now most of the models are suppressed with respect to $\Lambda$CDM, while  values of $w_p>-1$ (M2) enhance $D$. 

\begin{figure}[t!]
  \centering
  \includegraphics[scale=0.56]{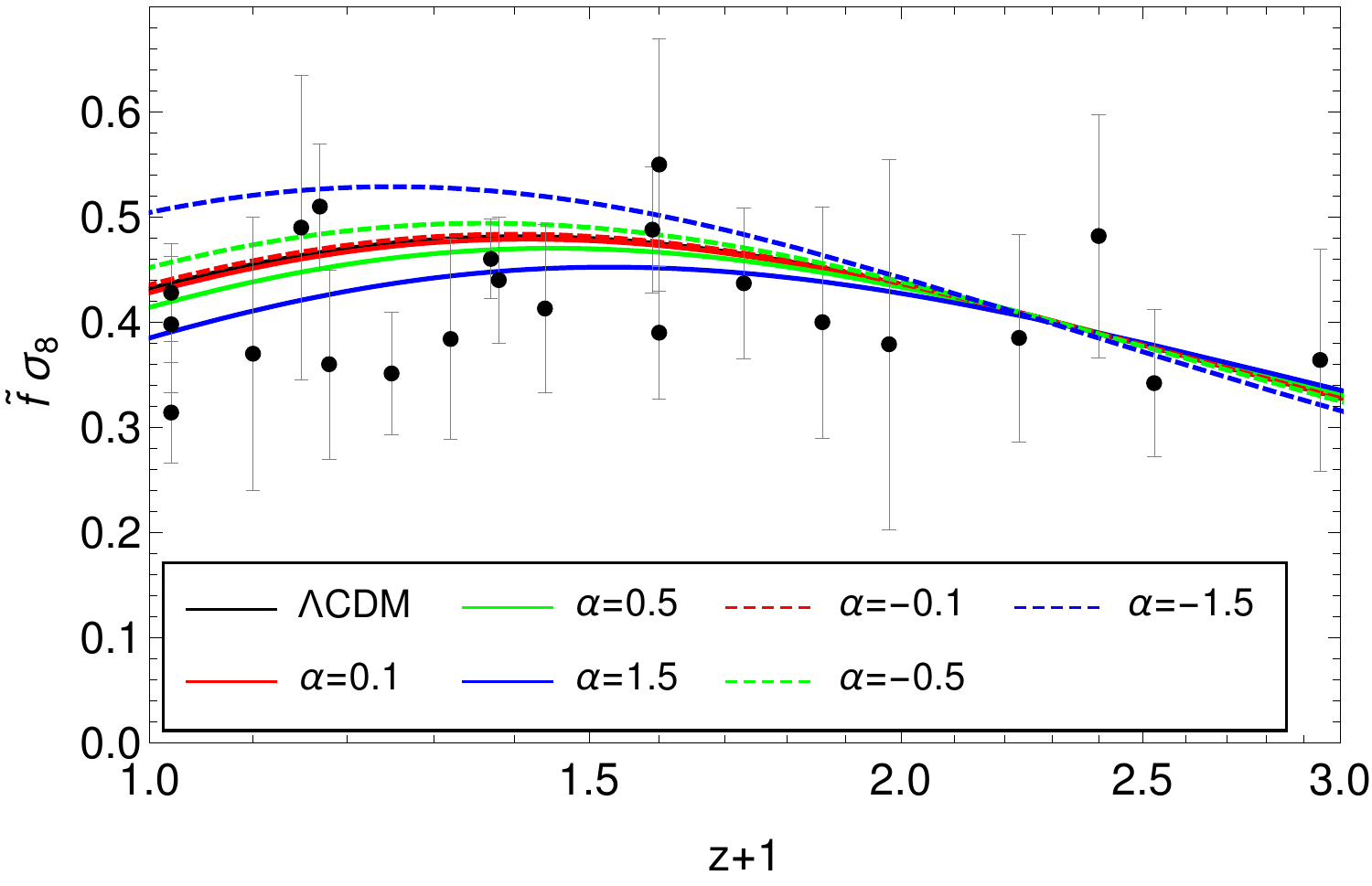}
\caption{Evolution of $\tilde{f}\sigma_8$  as a function of redshift $z$ for models with a $\Lambda$CDM background expansion history. We have used $\sigma_8^0=0.82$. The $\Lambda$CDM model  (solid, black line) is also included for comparison. Data set is from Ref. \cite{Sagredo:2018ahx}.}
\label{fig:fsigma8pltswQ0}
\end{figure}

\begin{figure}[t!]
  \centering
  \includegraphics[scale=0.56]{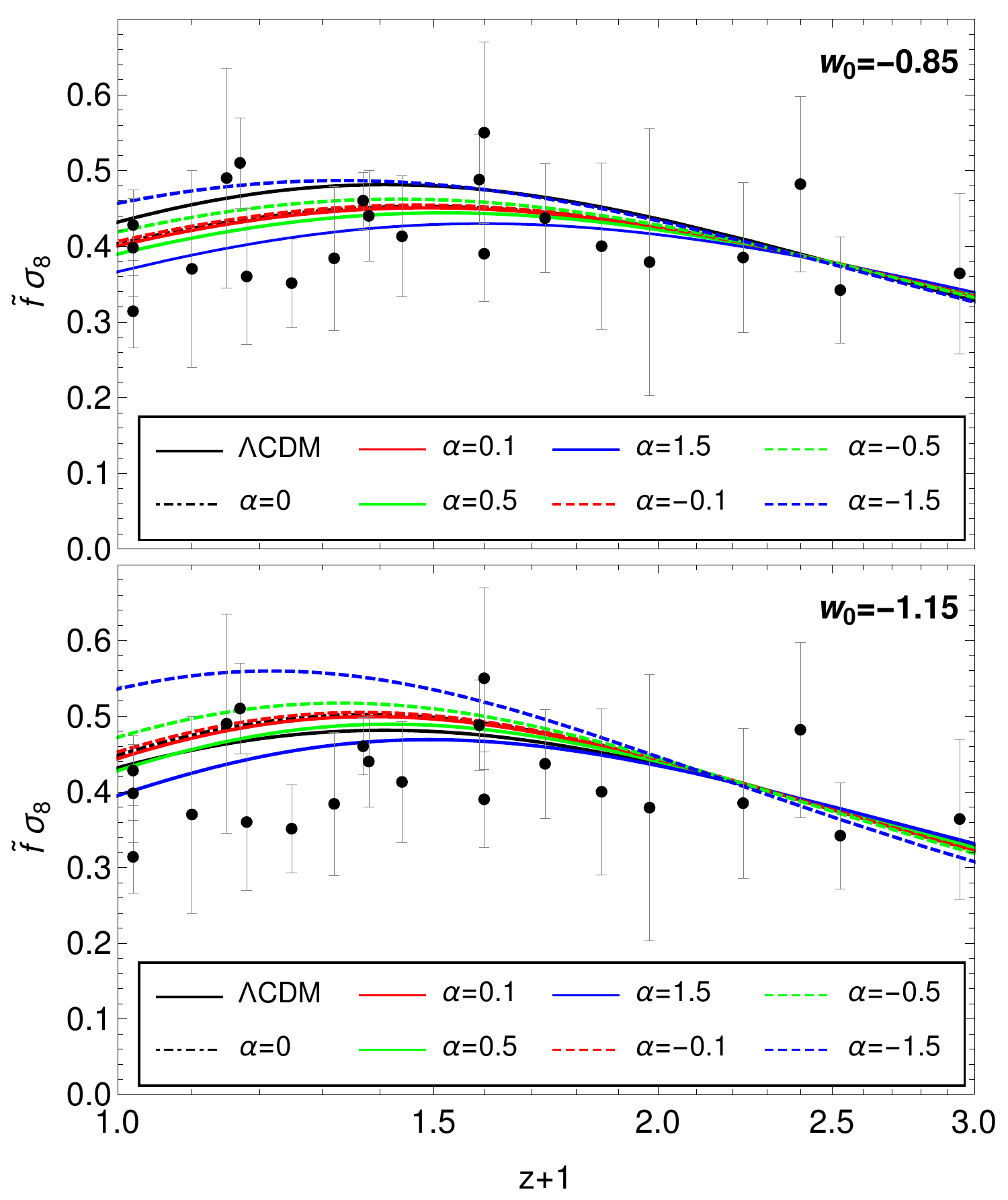}
\caption{Evolution of $\tilde{f}\sigma_8$  as a function of redshift $z$ for models with $w_0 = -0.85$ (top panel) and $w_0 = -1.15$ (bottom panel). We have used $\sigma_8^0=0.82$. The $\Lambda$CDM model  (solid, black line) is also included for comparison. Data set is from Ref. \cite{Sagredo:2018ahx}.} \label{fig:fsigma8pltsVWvwQ}
\end{figure}

\begin{figure}[t!]
  \centering
  \includegraphics[scale=0.56]{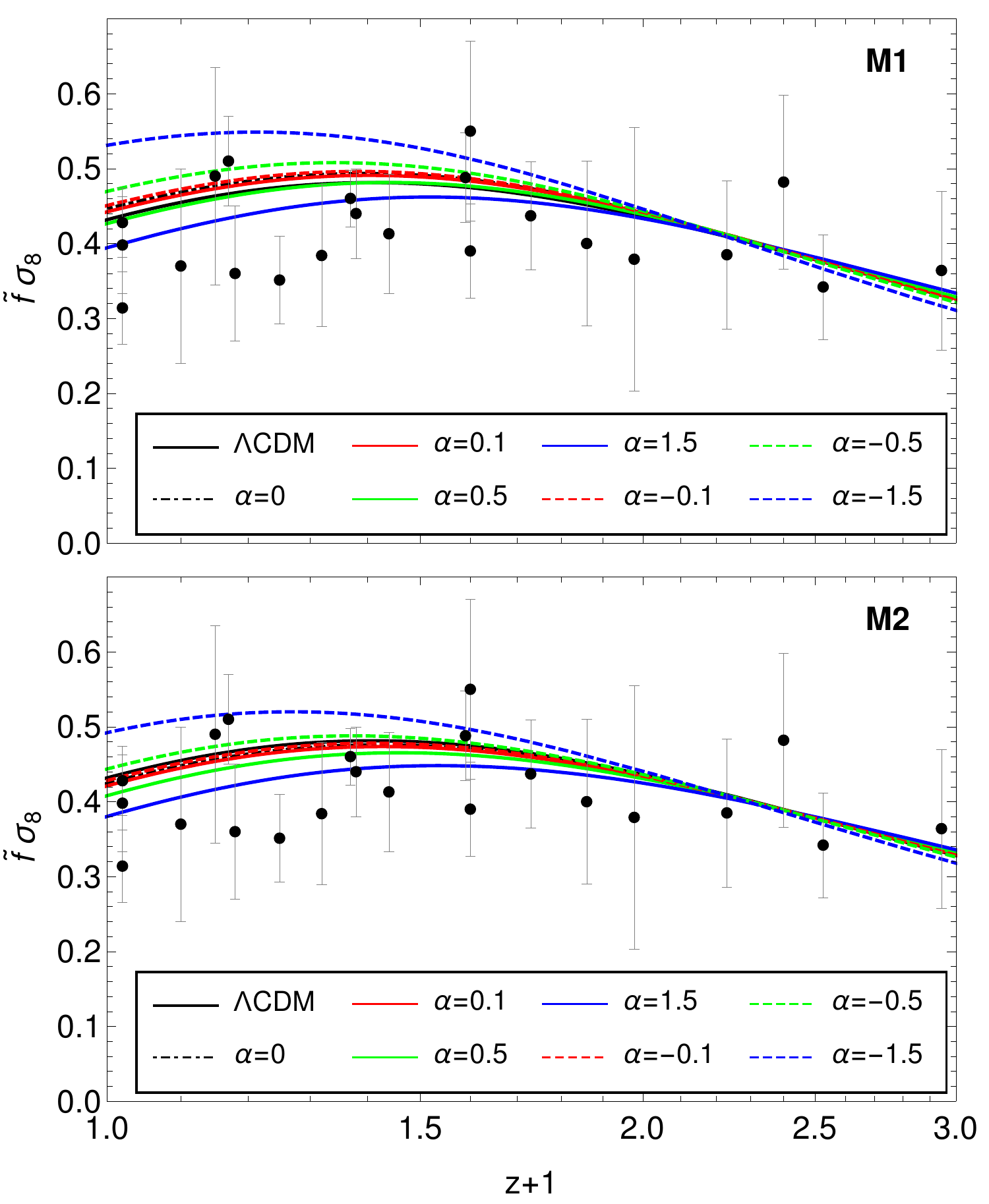}
\caption{Evolution of $\tilde{f}\sigma_8$  as a function of redshift $z$ for models with a time-dependent equation of state: M1 (top panel) and M2 (bottom panel). We have used $\sigma_8^0=0.82$. The $\Lambda$CDM model  (solid, black line) is also included for comparison. Data set is from Ref. \cite{Sagredo:2018ahx}.}
\label{fig:fsigma8pltsDwQ}
\end{figure}

We will now move to the analysis of a specific physical quantity that can be constructed from the linear growth factor, i.e. $\tilde{f}\sigma_8$, which is measured by redshift surveys and defined as the product of the growth rate,
$\tilde{f}\equiv \frac{d \ln \delta_{\rm m}}{d \ln a}$, and the root mean square of matter fluctuations, $\sigma_8$:
\begin{equation}
\tilde{f}\sigma_8=\sigma_8^0\frac{\delta^\prime(\ln a)}{\delta(\ln a=0)}\,,
\end{equation}
where $\sigma_8^0=\sigma_8(\ln a=0)$.

In Figures~\ref{fig:fsigma8pltswQ0}, \ref{fig:fsigma8pltsVWvwQ} and \ref{fig:fsigma8pltsDwQ} we plot the evolution of $\tilde{f}\sigma_8$ with $z$ for $0<z<2$ for the different cases under consideration. We also include RSD data from \cite{Sagredo:2018ahx}.
As expected from the discussion in Section~\ref{Sec:effectivecoupling}, the models with $\alpha<0$ prefer higher values for $\tilde{f}\sigma_8$, while data prefers a suppressed growth rate. It is also possible to adjust the amplitude by playing with the values of the effective equation of state. In this case, small and negative values of $\alpha$ can also show a lower amplitude for $\tilde{f}\sigma_8$, even though not comparable to the cases with $\alpha>0$. We can conclude that the most negative values of $\alpha$, in particular those associated either with a phantom behaviour for the effective equation of state or its time variation at early time, are unlikely to be preferred by RSD data.

In order to examine which values of $\alpha$ lead to a favored $f(Q)$ model over the $\Lambda$CDM one, it is necessary to perform a Markov-chain-Monte-Carlo simulation by varying $\sigma_8 (\ln a = 0)$ besides other cosmological parameters instead of fixing them. We leave this investigation for a future work.

\section{The ISW-galaxy cross-correlation} \label{Sec:ISWgal}

In this section we explore the sign of the cross-correlation between the ISW signal in CMB and galaxy distributions for the $f(Q)$ theory. $\Lambda$CDM and dark energy scenarios within GR have a positive ISW-galaxy cross-correlation  at any redshift, while for MG theories there can be cases for which the ISW-galaxy cross-correlation is negative. These cases will be ruled out by data \cite{Renk:2017rzu}. Therefore, this observable can be used as a tool to test viable models and distinguish between different cosmological scenarios \cite{Song:2006ej,Barreira:2012kk,Renk:2017rzu,Frusciante:2019puu,Giacomello:2018jfi,Hang:2021kfx,Kable:2021yws}. In a MG framework the modifications with respect to the standard scenario come from two sources: the  growth of structures and the late time ISW-effect. As discussed in the previous section, the growth of structures is strongly affected by the effective gravitational coupling, while the late-time ISW effect is sourced by the time derivative of the lensing potential and as such by the time derivative of the light deflection parameter. In $f(Q)$ gravity these two parameters are the same. Therefore, given the results of the previous sections we expect to gain significant information on the $\{ w_Q , \alpha \}$ parameter space by studying the sign of ISW-galaxy cross-correlation.

The ISW contribution to the CMB anisotropies can be written as
\begin{equation}
    \frac{\Delta T_{\rm ISW} (\mathbf{\hat{n}})}{T} = \int \frac{d \left( \Psi + \Phi \right)}{dz} dz \, .
\end{equation}
On the other hand, the fluctuations in the angular distribution of galaxies can be given by
\begin{equation}
    \frac{\Delta N_{\rm g} (\mathbf{\hat{n}}) }{N} = \int b_{\rm g} (k,z) \delta_{\rm m} (\mathbf{\hat{n}},z) \mathcal{W} (z) \, ,
\end{equation}
where $b_{\rm g}$ is the galaxy bias and $\mathcal{W}(z)$ is the selection function. As such, the cross-correlation power spectrum of the galaxy fluctuations and CMB anisotropies, written in the harmonic space, is
\begin{equation}
    \Bigg\langle \frac{\Delta T_{\rm ISW} (\mathbf{\hat{n}})}{T} \frac{\Delta N_{\rm g} (\mathbf{\hat{n}'})}{N} \Bigg\rangle = \sum_{\ell} \frac{2\ell + 1}{4\pi} \, C_{\ell} \, \mathcal{P}_{\ell} \left(\cos \theta \right) ,
\end{equation}
with $\mathcal{P}_{\ell}$ being the Legendre polynomial and $\theta$ the angle between the unit vectors $\mathbf{\hat{n}}$ and $\mathbf{\hat{n}'}$. Finally, $C_{\ell}$ is the amplitude of the ISW-galaxy cross-correlation which, upon the employment of the Limber approximation, can be written as~\cite{Kimura:2011td}
\begin{eqnarray}
   && C_{\ell} =  \frac{3 H_0^2 \, \Omega_{\rm m,0}}{\left( \ell + 1/2 \right)^2} \int dz \, H(z) \, \mathcal{W}(z) \, \frac{D}{D(z=0)} \nonumber \\
    && \times \frac{d U_{\rm ISW}}{dz} \, b_{\rm g} (k,z) \, P(k) \biggr\vert_{k=(\ell +1/2)/\chi} \, , 
\end{eqnarray}
where $P(k)$ is the present time matter power spectrum, $\chi$ is the comoving distance and
\begin{equation}
    U_{\rm ISW} (z) = \frac{ \Sigma \, D(z)}{D(z=0)a} \, .
    \label{eq:UISW}
\end{equation}
The latter has been defined considering  a scale independent growth which is precisely the case we are considering. This definition follows the lensing equation (\ref{eq:lenseq}) which  reads
\begin{equation}
    -k^2 \left( \Psi + \Phi \right) = 3 H_0^2 \, \Omega_{\rm m,0} \, U_{\rm ISW} (z) \, \delta_{\rm m} (z=0) \, .
\end{equation}

The amplitude of the ISW-galaxy cross-correlation is then proportional to the derivative 
\begin{equation}
\frac{dU_{\rm ISW}}{dz} = \frac{\Sigma D \mathcal{F}}{D(z=0)}\,,
\label{Eq:dUISW}
\end{equation}
where \cite{Nakamura:1811}
\begin{equation} \label{eq:fancyF}
    \mathcal{F} \equiv 1 - \frac{D'}{D} - \frac{\Sigma'}{\Sigma} \, .
\end{equation}
It follows that a necessary but not sufficient condition to have a negative ISW-galaxy cross-correlation for modified gravity models without a scale dependence in the effective couplings is \cite{Nakamura:1811}:
\begin{equation} \label{Eq:F}
    \mathcal{F} < 0 \, .
\end{equation}

Before presenting some results, let us discuss about  the regime of applicability of the  approach we use. First of all, we would like to stress that the expression in Eq.\ (\ref{Eq:dUISW})  does not rely on the quasi-static approximation because the way in which the Poisson and the lensing equations are parameterized in terms of $\mu$ and $\Sigma$ is general. The quasi-static approximation enters only when  an explicit and analytical expression for $\mu$ and $\Sigma$ is required (as it is in our case for $f(Q)$). Alternatively, one can solve the full equations and derive numerically the behaviour of $\mu$ and $\Sigma$ and finally place them back into Eq.\ (\ref{Eq:dUISW}). There is, nevertheless, an assumption to be considered, as we commented after Eq.\ (\ref{eq:UISW}). It has been assumed that $\delta_{\rm m}(z,k)$ depends only mildly on $k$, in which case it can be broken up in $k$ and time-dependent parts \cite{DeFelice:2011hq}. In our specific case, the analytic expression for $f(Q)$ derived within the quasi-static approximation is $k$-independent. Concerning the use of this approximation, the relevant scales for ISW data correspond to the modes deep inside the Hubble radius (namely, those for which $k^2/a^2 >> H^2$) which are in the regime of validity of the quasi-static approximation.

In the following we will provide theoretical predictions about the sign of the ISW-galaxy cross-correlation for $f(Q)$ gravity by analysing the sign of $\mathcal{F}$. In particular, this study will be informative in regards to the allowed sign and magnitude of $\alpha$. This is expected because dark energy models show a positive ISW-galaxy cross-correlation, meaning the impact of the background parameters is quite limited: they can only enhance or suppress the ISW-galaxy cross-correlation with respect to $\Lambda$CDM but its sign will always remain positive. This leaves $\alpha$ as the only parameter that can change the sign of this observable, offering a chance at constraining it.

\begin{figure}[t!]
\centering
\includegraphics[scale=0.56]{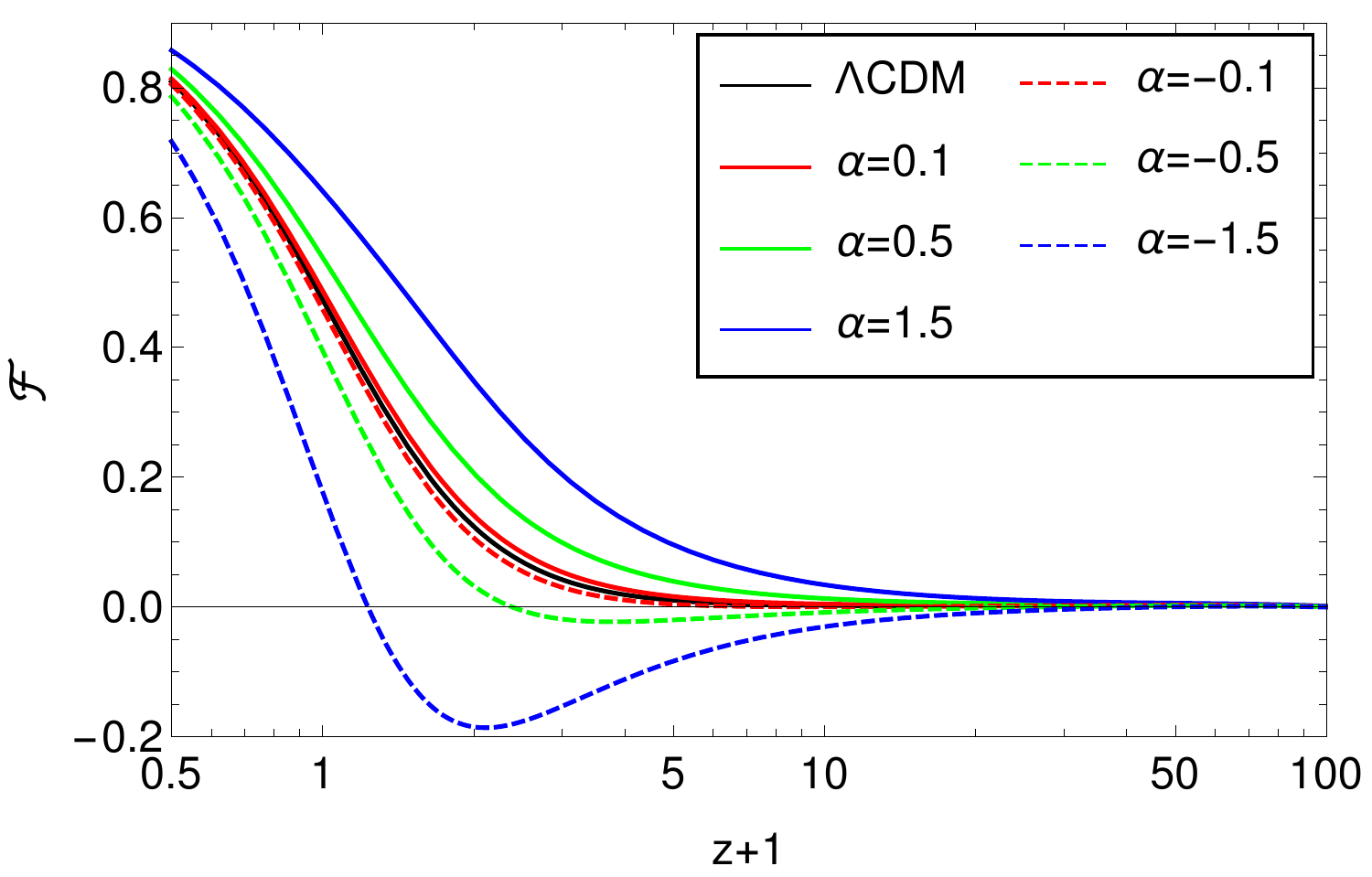}
\caption{Evolution of $\mathcal{F}$ as a function of redshift $z$ for models with an exact $\Lambda$CDM background evolution and different values of $\alpha$. \label{fig:Fpltsw-1}}
\end{figure}

In Figure~\ref{fig:Fpltsw-1} we show the evolution of $\mathcal{F}$ as a function of the redshift for the case with an exact $\Lambda$CDM background. We note that for all models with $\alpha$>0, $\mathcal{F}$ is higher than that of $\Lambda$CDM whereas for $\alpha$<0 all models fall bellow. Additionally, at present time we find that all values of $\alpha$ considered verify $\mathcal{F}(z=0)$>0. However, for negative values of $\alpha$, $\mathcal{F}$ can be negative at some redshift. In particular, the smaller and more negative the values of $\alpha$ are, the closer to present time in redshift $\mathcal{F}$ turns in the negative side. This aspect may set a bound on the negative branch of $\alpha$. This type of behaviour for $\mathcal{F}$ can be expected from the previously studied behaviours of $D$ and $\mu$, since $f(Q)$ verifies $\Sigma = \mu$. In fact, the term $D'/D$ of eq.~(\ref{eq:fancyF}) is positive during the entire evolution for all values of $\alpha$, meaning it will always have a negative contribution to $\mathcal{F}$. However, the contribution from $\Sigma ' / \Sigma$ will follow the previously discussed behaviour of $\mu$ and show a net division between positive and negative $\alpha$: for $\alpha$>0, $\Sigma' / \Sigma$ is negative at all $z$ while the opposite holds for $\alpha$<0. This means that in the cases with positive $\alpha$ it would only be possible to have $\mathcal{F}$<0 if $|D'/D|$>$1+\Sigma ' / \Sigma$. On the other hand, $\alpha$<0 models have two negative contributions for eq.~(\ref{eq:fancyF}), making it easier to reproduce a negative $\mathcal{F}$.

\begin{figure}[t!]
  \centering
  \includegraphics[scale=0.56]{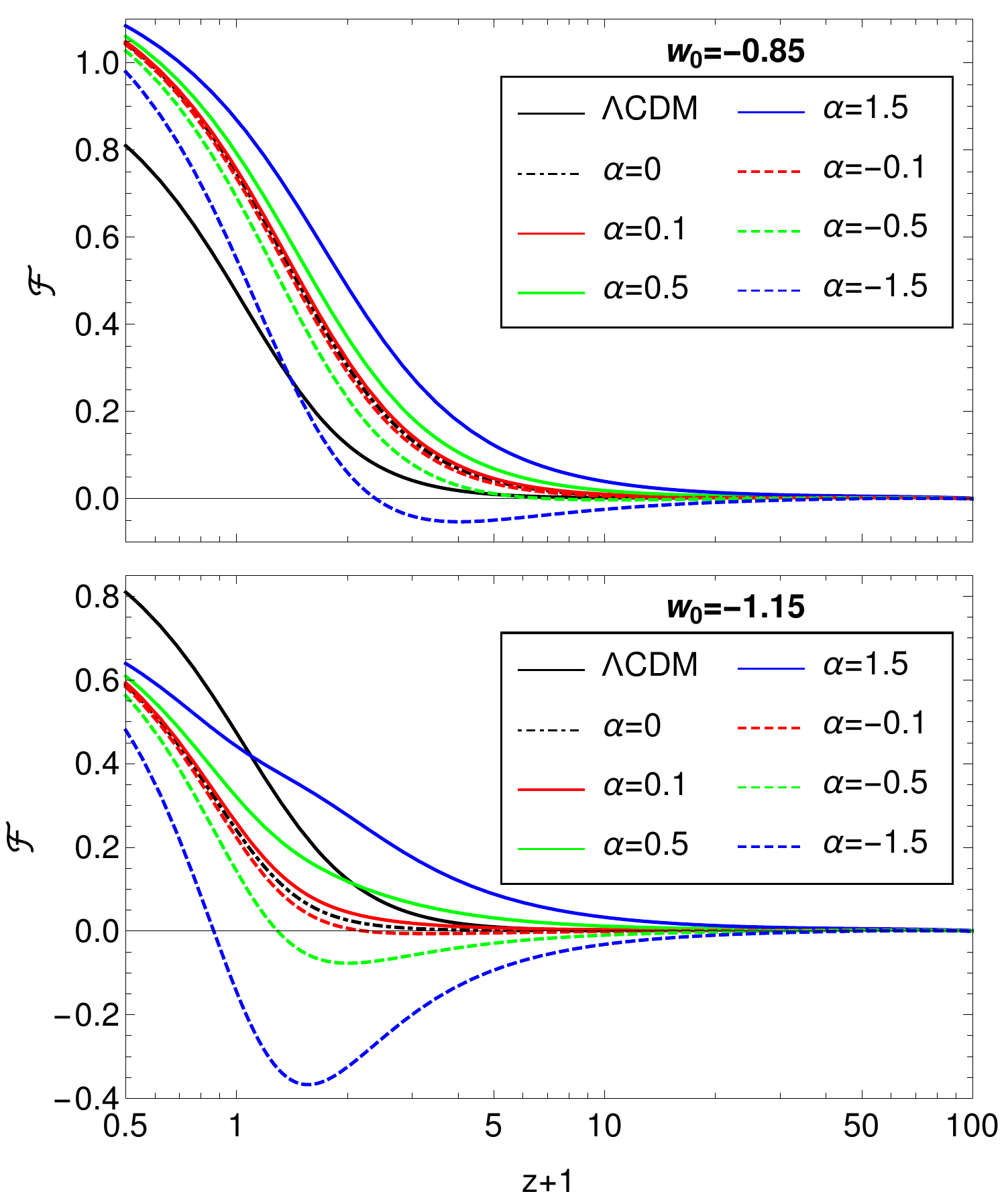}
  \caption{Evolution of $\mathcal{F}$ as a function of redshift $z$ for models with different values of $\alpha$ and fixed $w_0=-0.85$ (top panel) and $w_0 = -1.15$ (bottom panel). The $\Lambda$CDM case (solid, black line) is also included for comparison.}
\label{fig:FpltsVWvwQ}
\end{figure}

Then, in Figure~\ref{fig:FpltsVWvwQ} we show how the sign of $\mathcal{F}$ changes with redshift when the effective dark energy component has $w_Q=w_0$. From these figures it is clear the existent degeneracy between the value of $w_0$ and $\alpha$, which can be adjusted to have a positive $\mathcal{F}$ at all $z$ for the $f(Q)$ models. In details, when $w_0>-1$, 
we can notice that the model with $\alpha=0$ (black, dot-dashed line) is enhanced with respect to the $\Lambda$CDM case, thus favouring a positive $\mathcal{F}$ at all $z$. However, when $\alpha\neq0$ the amplitude may change and it is possible to find $\mathcal{F}$<0 when one has large negative values of $\alpha$. In this case, while at present time the sign of $\mathcal{F}$ is positive, at earlier time it can be negative. If we consider $w_0<-1$, the curve corresponding to $\alpha=0$ is suppressed with respect to $\Lambda$CDM. As such, in order to have $\mathcal{F}_{f(Q)}>\mathcal{F}_{\Lambda CDM}$ at all $z$, $\alpha$ has to be positive and large (e.g. $\alpha>0.5$). On the contrary, to avoid having $\mathcal{F}<0$, the smaller and negative values of $\alpha$ need to be excluded. 
As for the previous case, we find $D'/D$>0 for all values of $\alpha$. However, following the corresponding behaviour of $\mu$, there is no longer a net division between the effects of  positive and negative $\alpha$, which thus reflects on the evolution of $ \Sigma ' / \Sigma $. 
Models with the larger amplitude of the effective gravitational coupling  will accordingly show a larger and positive amplitude for  $\Sigma ' / \Sigma$ (specially at later time), consequently showing a negative $\mathcal{F}$. This is indeed the case for the more negative values of $\alpha$.

\begin{figure}[t!]
  \centering
  \includegraphics[scale=0.56]{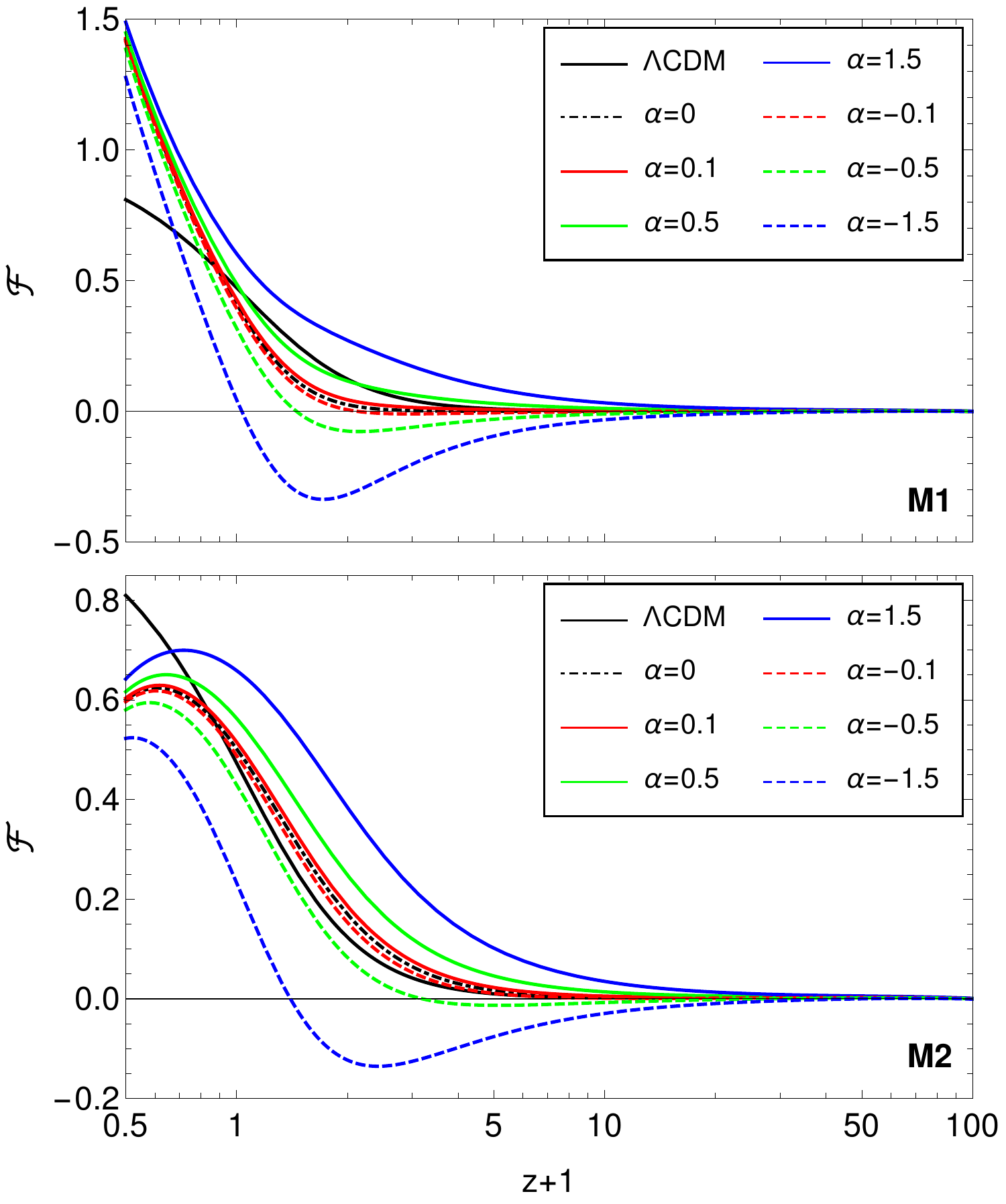}
\caption{Evolution of $\mathcal{F}$ as a function of redshift $z$ for the time-dependent equation of state: M1 (top panel) and M2 (bottom panel). The $\Lambda$CDM scenario (solid, black line) is also included for comparison.}
\label{fig:Fplts1DwQ}
\end{figure}

For the time-dependent background evolution we show the results in Figure~\ref{fig:Fplts1DwQ}. We note that, for cases with $w_p < w_0$ (phantom behaviour at early time, M1 model in the top panel) the $\mathcal{F}$ quantity is mostly suppressed with respect to $\Lambda$CDM, except for the largest value of $\alpha$. This is again connected to the balance of the two terms in eq.~(\ref{eq:fancyF}). Since $D'/D$ remains positive at all $z$, models with a larger $\mu$ and consequently a larger and positive $\Sigma'/\Sigma$ term will work to bring $\mathcal{F}$ to smaller amplitudes. Alternatively, models with smaller $\mu$ such as $\alpha=1.5$, are capable of having $\Sigma'/\Sigma$<0 which means only $D'/D$ contributes to the decrease of $\mathcal{F}$. In this case, we conclude that large  positive values of $\alpha$ will have $\mathcal{F}>0$ at all $z$, while in the other cases the sign of $\mathcal {F}$ depends on the magnitude of $|\alpha|$. In particular, large negative $|\alpha|$ have $\mathcal{F}<0$ at early time and then eventually cross the zero at later time. The crossing time depends on the $|\alpha|$: large values of $|\alpha|$ correspond to a crossing time very close to present day. 
For $w_p > w_0$ (M2, bottom panel), the models are enhanced with respect to $\Lambda$CDM favouring a positive ISW-galaxy cross-correlation, with the exception of the large negative $|\alpha|$. The situation is similar to the previous case. The $D'/D$ term is once again positive for all $z$, but this time the majority of the $\alpha$ have small amplitudes of the effective gravitational coupling, resulting in larger amplitudes for the $\mathcal{F}$ quantity.  

In summary, the analysis of the sign of $\mathcal{F}$ can give us an indication of a theoretical bound on the $\alpha$ parameter, which seems to exclude the most negative ones despite its degeneracy with $w_Q$. While here we have shown the evolution of $\mathcal{F}$ up to early times, datasets used to measure the cross-correlation of the CMB with tracers of the Large-Scale Structure (LSS) of the Universe are limited to low redshift ($z\simeq 1$). The latter include data from WISE, SDSS, SuperCOSMOS and NVSS.
Future surveys such as Euclid, LSST and SKA  may provide accurate data to measure this cross-correlation and set stringent bounds on the $f(Q)$ model and, more generally, to any dark energy and modified gravity model.

\section{Conclusion} \label{Sec:conclusion}

We have studied the dynamics of linear cosmological perturbations in the $f(Q)$ gravity model. Rather than fixing a specific functional form for $f(Q)$, we used the \textit{designer} approach, which allowed us to reconstruct the  form of the $f(Q)$ function associated to a chosen expansion history. We have introduced the parameter $\alpha$ which characterizes the family of solutions of $f(Q)$. This is one of the main results of this paper, being the first time the \textit{designer} approach is applied to $f(Q)$ gravity. To this extent, we have selected three effective equations of state, $w_Q$, for the effective dark energy component associated to $Q$, namely $w_Q=-1$, $w_Q=w_0$ and a fast varying equation of state. These choices are dictated not only by the aim of exploring different evolutions of $f(Q)$ models with redshift but also by the aim of studying a possible degeneracy between $\alpha$ and the parameters characterizing $w_Q$. 

We have provided theoretical predictions for a large set of linear phenomena:  the linear growth of structures, which included $\tilde{f}\sigma_8$, and the cross-correlation between the CMB and galaxy surveys. These physical quantities have one common denominator: the effective gravitational coupling, which is equal to the light deflection parameter  in the $f(Q)$ gravity. Deviations of these parameters from the standard scenario can then affect many observables and as such only a global study of the latter can allow us to draw some general conclusions.

For all observables investigated, we find that a different expansion history can lead to a solution for the $f(Q)$ behaviour that is able to impact them significantly. Let us summarize the main results in the following:
\begin{itemize}
    \item  We found a degeneracy between the $\alpha$ parameter and those of $w_Q$. In fact, we verified that in several occasions either $\alpha$ or the $w_Q$ parameters can be tuned in order to compensate or even reproduce the effect of the other. As such, one advantage that comes from using the \textit{designer} approach is that an appropriate choice of datasets used to perform parameter estimation with Monte Carlo Markov Chain methods can in principle break this degeneracy. Since $\alpha$ does not enter in the background expansion history, the degeneracy can be mitigated by using accurate background probes which strongly constrain $w_Q$, such that any effect at perturbations level needs to be modelled only with $\alpha$. We also stress that using GWs standard sirens may help in breaking the degeneracy as they can strongly constrain $f_Q$ \cite{Frusciante:2021sio}.
    \item We found that large negative values of $\alpha$, regardless of the chosen expansion history, can be excluded. These values indeed lead to a stronger gravity with respect to $\Lambda$CDM and as such to an higher amplitude for the $\tilde{f}\sigma_8$, which are unlikely to be favoured by RSD data only. Furthermore, models characterized by large negative values of $\alpha$ show a negative ISW-galaxy cross-correlation, which is excluded by observations.
\end{itemize}
     
Nevertheless, let us note that negative values of $\alpha$ can give rise to a lower ISW tail in the temperature-temperature power spectrum \cite{Frusciante:2021sio}, a feature that has been proved to be responsible for a better fit to CMB data for some MG models when compared to $\Lambda$CDM \cite{Peirone:2019aua,Frusciante:2019puu}, and which has recently been confirmed to be true for the case of a $f(Q)$ model with an exact $\Lambda$CDM background \cite{Atayde:2021ujc}. Therefore, while large negative values might be excluded this is not the case for small negative ones.

In conclusion, this type of phenomenological analysis of $f(Q)$ models provides insight on the types of deviations that might be expected on cosmological observables and that can be used to distinguish it from the $\Lambda$CDM scenario. Therefore it will be of interest to constrain the \textit{designer} $f(Q)$ gravity from the combined data analysis of LSS, CMB, BAO and SNIa.

\section*{Acknowledgements}

We thank B. J. Barros for useful comments on the manuscript. 
This work is  supported by Funda\c{c}\~{a}o para a  Ci\^{e}ncia e a Tecnologia (FCT) through the research grants UIDB/04434/2020, UIDP/04434/2020, PTDC/FIS-OUT\\/29048/2017, CERN/FIS-PAR/0037/2019 and by the FCT project ``BEYLA --BEYond LAmbda" with ref. number PTDC/FIS-AST/0054/2021. The research of ISA has received funding from the FCT PhD fellowship grant with ref.~number 2020.07237.BD. NF  acknowledges support from the personal FCT grant ``CosmoTests -- Cosmological tests of gravity theories beyond General Relativity" with ref.~number CEECIND/00017/2018.

\bibliography{Bib_entries}

\begin{thebibliography}{94}
\expandafter\ifx\csname natexlab\endcsname\relax\def\natexlab#1{#1}\fi
\providecommand{\url}[1]{\texttt{#1}}
\providecommand{\href}[2]{#2}
\providecommand{\path}[1]{#1}
\providecommand{\DOIprefix}{doi:}
\providecommand{\ArXivprefix}{arXiv:}
\providecommand{\URLprefix}{URL: }
\providecommand{\Pubmedprefix}{pmid:}
\providecommand{\doi}[1]{\href{http://dx.doi.org/#1}{\path{#1}}}
\providecommand{\Pubmed}[1]{\href{pmid:#1}{\path{#1}}}
\providecommand{\bibinfo}[2]{#2}
\ifx\xfnm\relax \def\xfnm[#1]{\unskip,\space#1}\fi
\bibitem[{Riess et~al.(1998)}]{SupernovaSearchTeam:1998fmf}
\bibinfo{author}{A.~G. Riess}, et~al. (\bibinfo{collaboration}{Supernova Search
  Team}),
\newblock \bibinfo{title}{{Observational evidence from supernovae for an
  accelerating universe and a cosmological constant}},
\newblock \bibinfo{journal}{Astron. J.} \bibinfo{volume}{116}
  (\bibinfo{year}{1998}) \bibinfo{pages}{1009--1038}.
  \DOIprefix\doi{10.1086/300499}.
  \href{http://arxiv.org/abs/astro-ph/9805201}{{\tt arXiv:astro-ph/9805201}}.
\bibitem[{Perlmutter et~al.(1999)}]{SupernovaCosmologyProject:1998vns}
\bibinfo{author}{S.~Perlmutter}, et~al. (\bibinfo{collaboration}{Supernova
  Cosmology Project}),
\newblock \bibinfo{title}{{Measurements of $\Omega$ and $\Lambda$ from 42 high
  redshift supernovae}},
\newblock \bibinfo{journal}{Astrophys. J.} \bibinfo{volume}{517}
  (\bibinfo{year}{1999}) \bibinfo{pages}{565--586}.
  \DOIprefix\doi{10.1086/307221}.
  \href{http://arxiv.org/abs/astro-ph/9812133}{{\tt arXiv:astro-ph/9812133}}.
\bibitem[{Spergel et~al.(2003)}]{WMAP:2003elm}
\bibinfo{author}{D.~N. Spergel}, et~al. (\bibinfo{collaboration}{WMAP}),
\newblock \bibinfo{title}{{First year Wilkinson Microwave Anisotropy Probe
  (WMAP) observations: Determination of cosmological parameters}},
\newblock \bibinfo{journal}{Astrophys. J. Suppl.} \bibinfo{volume}{148}
  (\bibinfo{year}{2003}) \bibinfo{pages}{175--194}.
  \DOIprefix\doi{10.1086/377226}.
  \href{http://arxiv.org/abs/astro-ph/0302209}{{\tt arXiv:astro-ph/0302209}}.
\bibitem[{Eisenstein et~al.(2005)}]{SDSS:2005xqv}
\bibinfo{author}{D.~J. Eisenstein}, et~al. (\bibinfo{collaboration}{SDSS}),
\newblock \bibinfo{title}{{Detection of the Baryon Acoustic Peak in the
  Large-Scale Correlation Function of SDSS Luminous Red Galaxies}},
\newblock \bibinfo{journal}{Astrophys. J.} \bibinfo{volume}{633}
  (\bibinfo{year}{2005}) \bibinfo{pages}{560--574}.
  \DOIprefix\doi{10.1086/466512}.
  \href{http://arxiv.org/abs/astro-ph/0501171}{{\tt arXiv:astro-ph/0501171}}.
\bibitem[{Betoule et~al.(2014)}]{SDSS:2014iwm}
\bibinfo{author}{M.~Betoule}, et~al. (\bibinfo{collaboration}{SDSS}),
\newblock \bibinfo{title}{{Improved cosmological constraints from a joint
  analysis of the SDSS-II and SNLS supernova samples}},
\newblock \bibinfo{journal}{Astron. Astrophys.} \bibinfo{volume}{568}
  (\bibinfo{year}{2014}) \bibinfo{pages}{A22}.
  \DOIprefix\doi{10.1051/0004-6361/201423413}.
  \href{http://arxiv.org/abs/1401.4064}{{\tt arXiv:1401.4064}}.
\bibitem[{Ade et~al.(2016)}]{Planck:2015fie}
\bibinfo{author}{P.~A.~R. Ade}, et~al. (\bibinfo{collaboration}{Planck}),
\newblock \bibinfo{title}{{Planck 2015 results. XIII. Cosmological
  parameters}},
\newblock \bibinfo{journal}{Astron. Astrophys.} \bibinfo{volume}{594}
  (\bibinfo{year}{2016}) \bibinfo{pages}{A13}.
  \DOIprefix\doi{10.1051/0004-6361/201525830}.
  \href{http://arxiv.org/abs/1502.01589}{{\tt arXiv:1502.01589}}.
\bibitem[{Aghanim et~al.(2016)}]{Planck:2015bpv}
\bibinfo{author}{N.~Aghanim}, et~al. (\bibinfo{collaboration}{Planck}),
\newblock \bibinfo{title}{{Planck 2015 results. XI. CMB power spectra,
  likelihoods, and robustness of parameters}},
\newblock \bibinfo{journal}{Astron. Astrophys.} \bibinfo{volume}{594}
  (\bibinfo{year}{2016}) \bibinfo{pages}{A11}.
  \DOIprefix\doi{10.1051/0004-6361/201526926}.
  \href{http://arxiv.org/abs/1507.02704}{{\tt arXiv:1507.02704}}.
\bibitem[{Weinberg(1989)}]{Weinberg:1988cp}
\bibinfo{author}{S.~Weinberg},
\newblock \bibinfo{title}{{The Cosmological Constant Problem}},
\newblock \bibinfo{journal}{Rev. Mod. Phys.} \bibinfo{volume}{61}
  (\bibinfo{year}{1989}) \bibinfo{pages}{1--23}.
  \DOIprefix\doi{10.1103/RevModPhys.61.1}.
\bibitem[{Carroll(2001)}]{Carroll:2000fy}
\bibinfo{author}{S.~M. Carroll},
\newblock \bibinfo{title}{{The Cosmological constant}},
\newblock \bibinfo{journal}{Living Rev. Rel.} \bibinfo{volume}{4}
  (\bibinfo{year}{2001}) \bibinfo{pages}{1}.
  \DOIprefix\doi{10.12942/lrr-2001-1}.
  \href{http://arxiv.org/abs/astro-ph/0004075}{{\tt arXiv:astro-ph/0004075}}.
\bibitem[{Velten et~al.(2014)Velten, vom Marttens, and
  Zimdahl}]{Velten:2014nra}
\bibinfo{author}{H.~E.~S. Velten}, \bibinfo{author}{R.~F. vom Marttens},
  \bibinfo{author}{W.~Zimdahl},
\newblock \bibinfo{title}{{Aspects of the cosmological
  \textquotedblleft{}coincidence problem\textquotedblright{}}},
\newblock \bibinfo{journal}{Eur. Phys. J. C} \bibinfo{volume}{74}
  (\bibinfo{year}{2014}) \bibinfo{pages}{3160}.
  \DOIprefix\doi{10.1140/epjc/s10052-014-3160-4}.
  \href{http://arxiv.org/abs/1410.2509}{{\tt arXiv:1410.2509}}.
\bibitem[{Joyce et~al.(2015)Joyce, Jain, Khoury, and Trodden}]{Joyce:2014kja}
\bibinfo{author}{A.~Joyce}, \bibinfo{author}{B.~Jain},
  \bibinfo{author}{J.~Khoury}, \bibinfo{author}{M.~Trodden},
\newblock \bibinfo{title}{{Beyond the Cosmological Standard Model}},
\newblock \bibinfo{journal}{Phys. Rept.} \bibinfo{volume}{568}
  (\bibinfo{year}{2015}) \bibinfo{pages}{1--98}.
  \DOIprefix\doi{10.1016/j.physrep.2014.12.002}.
  \href{http://arxiv.org/abs/1407.0059}{{\tt arXiv:1407.0059}}.
\bibitem[{Padilla(2015)}]{Padilla:2015aaa}
\bibinfo{author}{A.~Padilla},
\newblock \bibinfo{title}{{Lectures on the Cosmological Constant Problem}}
  (\bibinfo{year}{2015}). \href{http://arxiv.org/abs/1502.05296}{{\tt
  arXiv:1502.05296}}.
\bibitem[{Riess et~al.(2019)Riess, Casertano, Yuan, Macri, and
  Scolnic}]{Riess:2019cxk}
\bibinfo{author}{A.~G. Riess}, \bibinfo{author}{S.~Casertano},
  \bibinfo{author}{W.~Yuan}, \bibinfo{author}{L.~M. Macri},
  \bibinfo{author}{D.~Scolnic},
\newblock \bibinfo{title}{{Large Magellanic Cloud Cepheid Standards Provide a
  1\% Foundation for the Determination of the Hubble Constant and Stronger
  Evidence for Physics beyond $\Lambda$CDM}},
\newblock \bibinfo{journal}{Astrophys. J.} \bibinfo{volume}{876}
  (\bibinfo{year}{2019}) \bibinfo{pages}{85}.
  \DOIprefix\doi{10.3847/1538-4357/ab1422}.
  \href{http://arxiv.org/abs/1903.07603}{{\tt arXiv:1903.07603}}.
\bibitem[{Wong et~al.(2020)}]{Wong:2019kwg}
\bibinfo{author}{K.~C. Wong}, et~al.,
\newblock \bibinfo{title}{{H0LiCOW \textendash{} XIII. A 2.4 per cent
  measurement of H0 from lensed quasars: 5.3\ensuremath{\sigma} tension between
  early- and late-Universe probes}},
\newblock \bibinfo{journal}{Mon. Not. Roy. Astron. Soc.} \bibinfo{volume}{498}
  (\bibinfo{year}{2020}) \bibinfo{pages}{1420--1439}.
  \DOIprefix\doi{10.1093/mnras/stz3094}.
  \href{http://arxiv.org/abs/1907.04869}{{\tt arXiv:1907.04869}}.
\bibitem[{Delubac et~al.(2015)}]{BOSS:2014hwf}
\bibinfo{author}{T.~Delubac}, et~al. (\bibinfo{collaboration}{BOSS}),
\newblock \bibinfo{title}{{Baryon acoustic oscillations in the
  Ly\ensuremath{\alpha} forest of BOSS DR11 quasars}},
\newblock \bibinfo{journal}{Astron. Astrophys.} \bibinfo{volume}{574}
  (\bibinfo{year}{2015}) \bibinfo{pages}{A59}.
  \DOIprefix\doi{10.1051/0004-6361/201423969}.
  \href{http://arxiv.org/abs/1404.1801}{{\tt arXiv:1404.1801}}.
\bibitem[{Dawson et~al.(2013)}]{Dawson:2012va}
\bibinfo{author}{K.~S. Dawson}, et~al. (\bibinfo{collaboration}{BOSS}),
\newblock \bibinfo{title}{{The Baryon Oscillation Spectroscopic Survey of
  SDSS-III}},
\newblock \bibinfo{journal}{Astron. J.} \bibinfo{volume}{145}
  (\bibinfo{year}{2013}) \bibinfo{pages}{10}.
  \DOIprefix\doi{10.1088/0004-6256/145/1/10}.
  \href{http://arxiv.org/abs/1208.0022}{{\tt arXiv:1208.0022}}.
\bibitem[{Abazajian et~al.(2009)}]{SDSS:2008tqn}
\bibinfo{author}{K.~N. Abazajian}, et~al. (\bibinfo{collaboration}{SDSS}),
\newblock \bibinfo{title}{{The Seventh Data Release of the Sloan Digital Sky
  Survey}},
\newblock \bibinfo{journal}{Astrophys. J. Suppl.} \bibinfo{volume}{182}
  (\bibinfo{year}{2009}) \bibinfo{pages}{543--558}.
  \DOIprefix\doi{10.1088/0067-0049/182/2/543}.
  \href{http://arxiv.org/abs/0812.0649}{{\tt arXiv:0812.0649}}.
\bibitem[{Freedman et~al.(2019)}]{Freedman:2019jwv}
\bibinfo{author}{W.~L. Freedman}, et~al.,
\newblock \bibinfo{title}{{The Carnegie-Chicago Hubble Program. VIII. An
  Independent Determination of the Hubble Constant Based on the Tip of the Red
  Giant Branch}}  (\bibinfo{year}{2019}).
  \DOIprefix\doi{10.3847/1538-4357/ab2f73}.
  \href{http://arxiv.org/abs/1907.05922}{{\tt arXiv:1907.05922}}.
\bibitem[{Di~Valentino et~al.(2021)}]{DiValentino:2020zio}
\bibinfo{author}{E.~Di~Valentino}, et~al.,
\newblock \bibinfo{title}{{Snowmass2021 - Letter of interest cosmology
  intertwined II: The hubble constant tension}},
\newblock \bibinfo{journal}{Astropart. Phys.} \bibinfo{volume}{131}
  (\bibinfo{year}{2021}) \bibinfo{pages}{102605}.
  \DOIprefix\doi{10.1016/j.astropartphys.2021.102605}.
  \href{http://arxiv.org/abs/2008.11284}{{\tt arXiv:2008.11284}}.
\bibitem[{de~Jong et~al.(2015)}]{deJong:2015wca}
\bibinfo{author}{J.~T.~A. de~Jong}, et~al.,
\newblock \bibinfo{title}{{The first and second data releases of the
  Kilo-Degree Survey}},
\newblock \bibinfo{journal}{Astron. Astrophys.} \bibinfo{volume}{582}
  (\bibinfo{year}{2015}) \bibinfo{pages}{A62}.
  \DOIprefix\doi{10.1051/0004-6361/201526601}.
  \href{http://arxiv.org/abs/1507.00742}{{\tt arXiv:1507.00742}}.
\bibitem[{Hildebrandt et~al.(2017)}]{Hildebrandt:2016iqg}
\bibinfo{author}{H.~Hildebrandt}, et~al.,
\newblock \bibinfo{title}{{KiDS-450: Cosmological parameter constraints from
  tomographic weak gravitational lensing}},
\newblock \bibinfo{journal}{Mon. Not. Roy. Astron. Soc.} \bibinfo{volume}{465}
  (\bibinfo{year}{2017}) \bibinfo{pages}{1454}.
  \DOIprefix\doi{10.1093/mnras/stw2805}.
  \href{http://arxiv.org/abs/1606.05338}{{\tt arXiv:1606.05338}}.
\bibitem[{Kuijken et~al.(2015)}]{Kuijken:2015vca}
\bibinfo{author}{K.~Kuijken}, et~al.,
\newblock \bibinfo{title}{{Gravitational Lensing Analysis of the Kilo Degree
  Survey}},
\newblock \bibinfo{journal}{Mon. Not. Roy. Astron. Soc.} \bibinfo{volume}{454}
  (\bibinfo{year}{2015}) \bibinfo{pages}{3500--3532}.
  \DOIprefix\doi{10.1093/mnras/stv2140}.
  \href{http://arxiv.org/abs/1507.00738}{{\tt arXiv:1507.00738}}.
\bibitem[{Di~Valentino et~al.(2021)}]{DiValentino:2020vvd}
\bibinfo{author}{E.~Di~Valentino}, et~al.,
\newblock \bibinfo{title}{{Cosmology intertwined III: $f\sigma_8$ and $S_8$}},
\newblock \bibinfo{journal}{Astropart. Phys.} \bibinfo{volume}{131}
  (\bibinfo{year}{2021}) \bibinfo{pages}{102604}.
  \DOIprefix\doi{10.1016/j.astropartphys.2021.102604}.
  \href{http://arxiv.org/abs/2008.11285}{{\tt arXiv:2008.11285}}.
\bibitem[{Lue et~al.(2004)Lue, Scoccimarro, and Starkman}]{Lue:2004rj}
\bibinfo{author}{A.~Lue}, \bibinfo{author}{R.~Scoccimarro},
  \bibinfo{author}{G.~D. Starkman},
\newblock \bibinfo{title}{{Probing Newton's constant on vast scales: DGP
  gravity, cosmic acceleration and large scale structure}},
\newblock \bibinfo{journal}{Phys. Rev. D} \bibinfo{volume}{69}
  (\bibinfo{year}{2004}) \bibinfo{pages}{124015}.
  \DOIprefix\doi{10.1103/PhysRevD.69.124015}.
  \href{http://arxiv.org/abs/astro-ph/0401515}{{\tt arXiv:astro-ph/0401515}}.
\bibitem[{Copeland et~al.(2006)Copeland, Sami, and Tsujikawa}]{Copeland:2006wr}
\bibinfo{author}{E.~J. Copeland}, \bibinfo{author}{M.~Sami},
  \bibinfo{author}{S.~Tsujikawa},
\newblock \bibinfo{title}{{Dynamics of dark energy}},
\newblock \bibinfo{journal}{Int. J. Mod. Phys. D} \bibinfo{volume}{15}
  (\bibinfo{year}{2006}) \bibinfo{pages}{1753--1936}.
  \DOIprefix\doi{10.1142/S021827180600942X}.
  \href{http://arxiv.org/abs/hep-th/0603057}{{\tt arXiv:hep-th/0603057}}.
\bibitem[{Silvestri and Trodden(2009)}]{Silvestri:2009hh}
\bibinfo{author}{A.~Silvestri}, \bibinfo{author}{M.~Trodden},
\newblock \bibinfo{title}{{Approaches to Understanding Cosmic Acceleration}},
\newblock \bibinfo{journal}{Rept. Prog. Phys.} \bibinfo{volume}{72}
  (\bibinfo{year}{2009}) \bibinfo{pages}{096901}.
  \DOIprefix\doi{10.1088/0034-4885/72/9/096901}.
  \href{http://arxiv.org/abs/0904.0024}{{\tt arXiv:0904.0024}}.
\bibitem[{Nojiri and Odintsov(2011)}]{Nojiri:2010wj}
\bibinfo{author}{S.~Nojiri}, \bibinfo{author}{S.~D. Odintsov},
\newblock \bibinfo{title}{{Unified cosmic history in modified gravity: from
  F(R) theory to Lorentz non-invariant models}},
\newblock \bibinfo{journal}{Phys. Rept.} \bibinfo{volume}{505}
  (\bibinfo{year}{2011}) \bibinfo{pages}{59--144}.
  \DOIprefix\doi{10.1016/j.physrep.2011.04.001}.
  \href{http://arxiv.org/abs/1011.0544}{{\tt arXiv:1011.0544}}.
\bibitem[{Tsujikawa(2010)}]{Tsujikawa:2010zza}
\bibinfo{author}{S.~Tsujikawa},
\newblock \bibinfo{title}{{Modified gravity models of dark energy}},
\newblock \bibinfo{journal}{Lect. Notes Phys.} \bibinfo{volume}{800}
  (\bibinfo{year}{2010}) \bibinfo{pages}{99--145}.
  \DOIprefix\doi{10.1007/978-3-642-10598-2\_3}.
  \href{http://arxiv.org/abs/1101.0191}{{\tt arXiv:1101.0191}}.
\bibitem[{Capozziello and De~Laurentis(2011)}]{Capozziello:2011et}
\bibinfo{author}{S.~Capozziello}, \bibinfo{author}{M.~De~Laurentis},
\newblock \bibinfo{title}{{Extended Theories of Gravity}},
\newblock \bibinfo{journal}{Phys. Rept.} \bibinfo{volume}{509}
  (\bibinfo{year}{2011}) \bibinfo{pages}{167--321}.
  \DOIprefix\doi{10.1016/j.physrep.2011.09.003}.
  \href{http://arxiv.org/abs/1108.6266}{{\tt arXiv:1108.6266}}.
\bibitem[{Clifton et~al.(2012)Clifton, Ferreira, Padilla, and
  Skordis}]{Clifton:2011jh}
\bibinfo{author}{T.~Clifton}, \bibinfo{author}{P.~G. Ferreira},
  \bibinfo{author}{A.~Padilla}, \bibinfo{author}{C.~Skordis},
\newblock \bibinfo{title}{{Modified Gravity and Cosmology}},
\newblock \bibinfo{journal}{Phys. Rept.} \bibinfo{volume}{513}
  (\bibinfo{year}{2012}) \bibinfo{pages}{1--189}.
  \DOIprefix\doi{10.1016/j.physrep.2012.01.001}.
  \href{http://arxiv.org/abs/1106.2476}{{\tt arXiv:1106.2476}}.
\bibitem[{Bamba et~al.(2012)Bamba, Capozziello, Nojiri, and
  Odintsov}]{Bamba:2012cp}
\bibinfo{author}{K.~Bamba}, \bibinfo{author}{S.~Capozziello},
  \bibinfo{author}{S.~Nojiri}, \bibinfo{author}{S.~D. Odintsov},
\newblock \bibinfo{title}{{Dark energy cosmology: the equivalent description
  via different theoretical models and cosmography tests}},
\newblock \bibinfo{journal}{Astrophys. Space Sci.} \bibinfo{volume}{342}
  (\bibinfo{year}{2012}) \bibinfo{pages}{155--228}.
  \DOIprefix\doi{10.1007/s10509-012-1181-8}.
  \href{http://arxiv.org/abs/1205.3421}{{\tt arXiv:1205.3421}}.
\bibitem[{Koyama(2016)}]{Koyama:2015vza}
\bibinfo{author}{K.~Koyama},
\newblock \bibinfo{title}{{Cosmological Tests of Modified Gravity}},
\newblock \bibinfo{journal}{Rept. Prog. Phys.} \bibinfo{volume}{79}
  (\bibinfo{year}{2016}) \bibinfo{pages}{046902}.
  \DOIprefix\doi{10.1088/0034-4885/79/4/046902}.
  \href{http://arxiv.org/abs/1504.04623}{{\tt arXiv:1504.04623}}.
\bibitem[{Avelino et~al.(2016)}]{Avelino:2016lpj}
\bibinfo{author}{P.~Avelino}, et~al.,
\newblock \bibinfo{title}{{Unveiling the Dynamics of the Universe}},
\newblock \bibinfo{journal}{Symmetry} \bibinfo{volume}{8}
  (\bibinfo{year}{2016}) \bibinfo{pages}{70}.
  \DOIprefix\doi{10.3390/sym8080070}.
  \href{http://arxiv.org/abs/1607.02979}{{\tt arXiv:1607.02979}}.
\bibitem[{Joyce et~al.(2016)Joyce, Lombriser, and Schmidt}]{Joyce:2016vqv}
\bibinfo{author}{A.~Joyce}, \bibinfo{author}{L.~Lombriser},
  \bibinfo{author}{F.~Schmidt},
\newblock \bibinfo{title}{{Dark Energy Versus Modified Gravity}},
\newblock \bibinfo{journal}{Ann. Rev. Nucl. Part. Sci.} \bibinfo{volume}{66}
  (\bibinfo{year}{2016}) \bibinfo{pages}{95--122}.
  \DOIprefix\doi{10.1146/annurev-nucl-102115-044553}.
  \href{http://arxiv.org/abs/1601.06133}{{\tt arXiv:1601.06133}}.
\bibitem[{Nojiri et~al.(2017)Nojiri, Odintsov, and Oikonomou}]{Nojiri:2017ncd}
\bibinfo{author}{S.~Nojiri}, \bibinfo{author}{S.~D. Odintsov},
  \bibinfo{author}{V.~K. Oikonomou},
\newblock \bibinfo{title}{{Modified Gravity Theories on a Nutshell: Inflation,
  Bounce and Late-time Evolution}},
\newblock \bibinfo{journal}{Phys. Rept.} \bibinfo{volume}{692}
  (\bibinfo{year}{2017}) \bibinfo{pages}{1--104}.
  \DOIprefix\doi{10.1016/j.physrep.2017.06.001}.
  \href{http://arxiv.org/abs/1705.11098}{{\tt arXiv:1705.11098}}.
\bibitem[{Ferreira(2019)}]{Ferreira:2019xrr}
\bibinfo{author}{P.~G. Ferreira},
\newblock \bibinfo{title}{{Cosmological Tests of Gravity}},
\newblock \bibinfo{journal}{Ann. Rev. Astron. Astrophys.} \bibinfo{volume}{57}
  (\bibinfo{year}{2019}) \bibinfo{pages}{335--374}.
  \DOIprefix\doi{10.1146/annurev-astro-091918-104423}.
  \href{http://arxiv.org/abs/1902.10503}{{\tt arXiv:1902.10503}}.
\bibitem[{Kobayashi(2019)}]{Kobayashi:2019hrl}
\bibinfo{author}{T.~Kobayashi},
\newblock \bibinfo{title}{{Horndeski theory and beyond: a review}},
\newblock \bibinfo{journal}{Rept. Prog. Phys.} \bibinfo{volume}{82}
  (\bibinfo{year}{2019}) \bibinfo{pages}{086901}.
  \DOIprefix\doi{10.1088/1361-6633/ab2429}.
  \href{http://arxiv.org/abs/1901.07183}{{\tt arXiv:1901.07183}}.
\bibitem[{Frusciante and Perenon(2020)}]{Frusciante:2019xia}
\bibinfo{author}{N.~Frusciante}, \bibinfo{author}{L.~Perenon},
\newblock \bibinfo{title}{{Effective field theory of dark energy: A review}},
\newblock \bibinfo{journal}{Phys. Rept.} \bibinfo{volume}{857}
  (\bibinfo{year}{2020}) \bibinfo{pages}{1--63}.
  \DOIprefix\doi{10.1016/j.physrep.2020.02.004}.
  \href{http://arxiv.org/abs/1907.03150}{{\tt arXiv:1907.03150}}.
\bibitem[{Saridakis et~al.(2021)}]{CANTATA:2021ktz}
\bibinfo{author}{E.~N. Saridakis}, et~al. (\bibinfo{collaboration}{CANTATA}),
\newblock \bibinfo{title}{{Modified Gravity and Cosmology: An Update by the
  CANTATA Network}}  (\bibinfo{year}{2021}).
  \href{http://arxiv.org/abs/2105.12582}{{\tt arXiv:2105.12582}}.
\bibitem[{Bahamonde et~al.(2021)Bahamonde, Dialektopoulos, Escamilla-Rivera,
  Farrugia, Gakis, Hendry, Hohmann, Said, Mifsud, and
  Di~Valentino}]{Bahamonde:2021gfp}
\bibinfo{author}{S.~Bahamonde}, \bibinfo{author}{K.~F. Dialektopoulos},
  \bibinfo{author}{C.~Escamilla-Rivera}, \bibinfo{author}{G.~Farrugia},
  \bibinfo{author}{V.~Gakis}, \bibinfo{author}{M.~Hendry},
  \bibinfo{author}{M.~Hohmann}, \bibinfo{author}{J.~L. Said},
  \bibinfo{author}{J.~Mifsud}, \bibinfo{author}{E.~Di~Valentino},
\newblock \bibinfo{title}{{Teleparallel Gravity: From Theory to Cosmology}}
  (\bibinfo{year}{2021}). \href{http://arxiv.org/abs/2106.13793}{{\tt
  arXiv:2106.13793}}.
\bibitem[{Nunes(2018)}]{Nunes:2018xbm}
\bibinfo{author}{R.~C. Nunes},
\newblock \bibinfo{title}{{Structure formation in $f(T)$ gravity and a solution
  for $H_0$ tension}},
\newblock \bibinfo{journal}{JCAP} \bibinfo{volume}{05} (\bibinfo{year}{2018})
  \bibinfo{pages}{052}. \DOIprefix\doi{10.1088/1475-7516/2018/05/052}.
  \href{http://arxiv.org/abs/1802.02281}{{\tt arXiv:1802.02281}}.
\bibitem[{Zumalacarregui(2020)}]{Zumalacarregui:2020cjh}
\bibinfo{author}{M.~Zumalacarregui},
\newblock \bibinfo{title}{{Gravity in the Era of Equality: Towards solutions to
  the Hubble problem without fine-tuned initial conditions}},
\newblock \bibinfo{journal}{Phys. Rev. D} \bibinfo{volume}{102}
  (\bibinfo{year}{2020}) \bibinfo{pages}{023523}.
  \DOIprefix\doi{10.1103/PhysRevD.102.023523}.
  \href{http://arxiv.org/abs/2003.06396}{{\tt arXiv:2003.06396}}.
\bibitem[{Belgacem et~al.(2018)Belgacem, Dirian, Foffa, and
  Maggiore}]{Belgacem:2017cqo}
\bibinfo{author}{E.~Belgacem}, \bibinfo{author}{Y.~Dirian},
  \bibinfo{author}{S.~Foffa}, \bibinfo{author}{M.~Maggiore},
\newblock \bibinfo{title}{{Nonlocal gravity. Conceptual aspects and
  cosmological predictions}},
\newblock \bibinfo{journal}{JCAP} \bibinfo{volume}{03} (\bibinfo{year}{2018})
  \bibinfo{pages}{002}. \DOIprefix\doi{10.1088/1475-7516/2018/03/002}.
  \href{http://arxiv.org/abs/1712.07066}{{\tt arXiv:1712.07066}}.
\bibitem[{Rossi et~al.(2019)Rossi, Ballardini, Braglia, Finelli, Paoletti,
  Starobinsky, and Umilt\`a}]{Rossi:2019lgt}
\bibinfo{author}{M.~Rossi}, \bibinfo{author}{M.~Ballardini},
  \bibinfo{author}{M.~Braglia}, \bibinfo{author}{F.~Finelli},
  \bibinfo{author}{D.~Paoletti}, \bibinfo{author}{A.~A. Starobinsky},
  \bibinfo{author}{C.~Umilt\`a},
\newblock \bibinfo{title}{{Cosmological constraints on post-Newtonian
  parameters in effectively massless scalar-tensor theories of gravity}},
\newblock \bibinfo{journal}{Phys. Rev. D} \bibinfo{volume}{100}
  (\bibinfo{year}{2019}) \bibinfo{pages}{103524}.
  \DOIprefix\doi{10.1103/PhysRevD.100.103524}.
  \href{http://arxiv.org/abs/1906.10218}{{\tt arXiv:1906.10218}}.
\bibitem[{Peirone et~al.(2019)Peirone, Benevento, Frusciante, and
  Tsujikawa}]{Peirone:2019aua}
\bibinfo{author}{S.~Peirone}, \bibinfo{author}{G.~Benevento},
  \bibinfo{author}{N.~Frusciante}, \bibinfo{author}{S.~Tsujikawa},
\newblock \bibinfo{title}{{Cosmological data favor Galileon ghost condensate
  over $\Lambda$CDM}},
\newblock \bibinfo{journal}{Phys. Rev. D} \bibinfo{volume}{100}
  (\bibinfo{year}{2019}) \bibinfo{pages}{063540}.
  \DOIprefix\doi{10.1103/PhysRevD.100.063540}.
  \href{http://arxiv.org/abs/1905.05166}{{\tt arXiv:1905.05166}}.
\bibitem[{Frusciante et~al.(2020)Frusciante, Peirone, Atayde, and
  De~Felice}]{Frusciante:2019puu}
\bibinfo{author}{N.~Frusciante}, \bibinfo{author}{S.~Peirone},
  \bibinfo{author}{L.~Atayde}, \bibinfo{author}{A.~De~Felice},
\newblock \bibinfo{title}{{Phenomenology of the generalized cubic covariant
  Galileon model and cosmological bounds}},
\newblock \bibinfo{journal}{Phys. Rev. D} \bibinfo{volume}{101}
  (\bibinfo{year}{2020}) \bibinfo{pages}{064001}.
  \DOIprefix\doi{10.1103/PhysRevD.101.064001}.
  \href{http://arxiv.org/abs/1912.07586}{{\tt arXiv:1912.07586}}.
\bibitem[{Heisenberg and Villarrubia-Rojo(2021)}]{Heisenberg:2020xak}
\bibinfo{author}{L.~Heisenberg}, \bibinfo{author}{H.~Villarrubia-Rojo},
\newblock \bibinfo{title}{{Proca in the sky}},
\newblock \bibinfo{journal}{JCAP} \bibinfo{volume}{03} (\bibinfo{year}{2021})
  \bibinfo{pages}{032}. \DOIprefix\doi{10.1088/1475-7516/2021/03/032}.
  \href{http://arxiv.org/abs/2010.00513}{{\tt arXiv:2010.00513}}.
\bibitem[{Barros et~al.(2020)Barros, Barreiro, Koivisto, and
  Nunes}]{Barros:2020bgg}
\bibinfo{author}{B.~J. Barros}, \bibinfo{author}{T.~Barreiro},
  \bibinfo{author}{T.~Koivisto}, \bibinfo{author}{N.~J. Nunes},
\newblock \bibinfo{title}{{Testing $F(Q)$ gravity with redshift space
  distortions}},
\newblock \bibinfo{journal}{Phys. Dark Univ.} \bibinfo{volume}{30}
  (\bibinfo{year}{2020}) \bibinfo{pages}{100616}.
  \DOIprefix\doi{10.1016/j.dark.2020.100616}.
  \href{http://arxiv.org/abs/2004.07867}{{\tt arXiv:2004.07867}}.
\bibitem[{Barros et~al.(2019)Barros, Amendola, Barreiro, and
  Nunes}]{Barros:2018efl}
\bibinfo{author}{B.~J. Barros}, \bibinfo{author}{L.~Amendola},
  \bibinfo{author}{T.~Barreiro}, \bibinfo{author}{N.~J. Nunes},
\newblock \bibinfo{title}{{Coupled quintessence with a $\Lambda$CDM background:
  removing the $\sigma_8$ tension}},
\newblock \bibinfo{journal}{JCAP} \bibinfo{volume}{01} (\bibinfo{year}{2019})
  \bibinfo{pages}{007}. \DOIprefix\doi{10.1088/1475-7516/2019/01/007}.
  \href{http://arxiv.org/abs/1802.09216}{{\tt arXiv:1802.09216}}.
\bibitem[{Di~Valentino et~al.(2021)Di~Valentino, Mena, Pan, Visinelli, Yang,
  Melchiorri, Mota, Riess, and Silk}]{DiValentino:2021izs}
\bibinfo{author}{E.~Di~Valentino}, \bibinfo{author}{O.~Mena},
  \bibinfo{author}{S.~Pan}, \bibinfo{author}{L.~Visinelli},
  \bibinfo{author}{W.~Yang}, \bibinfo{author}{A.~Melchiorri},
  \bibinfo{author}{D.~F. Mota}, \bibinfo{author}{A.~G. Riess},
  \bibinfo{author}{J.~Silk},
\newblock \bibinfo{title}{{In the Realm of the Hubble tension $-$ a Review of
  Solutions}}  (\bibinfo{year}{2021}).
  \DOIprefix\doi{10.1088/1361-6382/ac086d}.
  \href{http://arxiv.org/abs/2103.01183}{{\tt arXiv:2103.01183}}.
\bibitem[{Beltr\'an~Jim\'enez et~al.(2018)Beltr\'an~Jim\'enez, Heisenberg, and
  Koivisto}]{BeltranJimenez:2017tkd}
\bibinfo{author}{J.~Beltr\'an~Jim\'enez}, \bibinfo{author}{L.~Heisenberg},
  \bibinfo{author}{T.~Koivisto},
\newblock \bibinfo{title}{{Coincident General Relativity}},
\newblock \bibinfo{journal}{Phys. Rev. D} \bibinfo{volume}{98}
  (\bibinfo{year}{2018}) \bibinfo{pages}{044048}.
  \DOIprefix\doi{10.1103/PhysRevD.98.044048}.
  \href{http://arxiv.org/abs/1710.03116}{{\tt arXiv:1710.03116}}.
\bibitem[{Harko et~al.(2018)Harko, Koivisto, Lobo, Olmo, and
  Rubiera-Garcia}]{Harko:2018gxr}
\bibinfo{author}{T.~Harko}, \bibinfo{author}{T.~S. Koivisto},
  \bibinfo{author}{F.~S.~N. Lobo}, \bibinfo{author}{G.~J. Olmo},
  \bibinfo{author}{D.~Rubiera-Garcia},
\newblock \bibinfo{title}{{Coupling matter in modified $Q$ gravity}},
\newblock \bibinfo{journal}{Phys. Rev. D} \bibinfo{volume}{98}
  (\bibinfo{year}{2018}) \bibinfo{pages}{084043}.
  \DOIprefix\doi{10.1103/PhysRevD.98.084043}.
  \href{http://arxiv.org/abs/1806.10437}{{\tt arXiv:1806.10437}}.
\bibitem[{Xu et~al.(2019)Xu, Li, Harko, and Liang}]{Xu:2019sbp}
\bibinfo{author}{Y.~Xu}, \bibinfo{author}{G.~Li}, \bibinfo{author}{T.~Harko},
  \bibinfo{author}{S.-D. Liang},
\newblock \bibinfo{title}{{$f(Q,T)$ gravity}},
\newblock \bibinfo{journal}{Eur. Phys. J. C} \bibinfo{volume}{79}
  (\bibinfo{year}{2019}) \bibinfo{pages}{708}.
  \DOIprefix\doi{10.1140/epjc/s10052-019-7207-4}.
  \href{http://arxiv.org/abs/1908.04760}{{\tt arXiv:1908.04760}}.
\bibitem[{J\"arv et~al.(2018)J\"arv, R\"unkla, Saal, and Vilson}]{Jarv:2018bgs}
\bibinfo{author}{L.~J\"arv}, \bibinfo{author}{M.~R\"unkla},
  \bibinfo{author}{M.~Saal}, \bibinfo{author}{O.~Vilson},
\newblock \bibinfo{title}{{Nonmetricity formulation of general relativity and
  its scalar-tensor extension}},
\newblock \bibinfo{journal}{Phys. Rev. D} \bibinfo{volume}{97}
  (\bibinfo{year}{2018}) \bibinfo{pages}{124025}.
  \DOIprefix\doi{10.1103/PhysRevD.97.124025}.
  \href{http://arxiv.org/abs/1802.00492}{{\tt arXiv:1802.00492}}.
\bibitem[{R\"unkla and Vilson(2018)}]{Runkla:2018xrv}
\bibinfo{author}{M.~R\"unkla}, \bibinfo{author}{O.~Vilson},
\newblock \bibinfo{title}{{Family of scalar-nonmetricity theories of gravity}},
\newblock \bibinfo{journal}{Phys. Rev. D} \bibinfo{volume}{98}
  (\bibinfo{year}{2018}) \bibinfo{pages}{084034}.
  \DOIprefix\doi{10.1103/PhysRevD.98.084034}.
  \href{http://arxiv.org/abs/1805.12197}{{\tt arXiv:1805.12197}}.
\bibitem[{Nester and Yo(1999)}]{Nester:1998mp}
\bibinfo{author}{J.~M. Nester}, \bibinfo{author}{H.-J. Yo},
\newblock \bibinfo{title}{{Symmetric teleparallel general relativity}},
\newblock \bibinfo{journal}{Chin. J. Phys.} \bibinfo{volume}{37}
  (\bibinfo{year}{1999}) \bibinfo{pages}{113}.
  \href{http://arxiv.org/abs/gr-qc/9809049}{{\tt arXiv:gr-qc/9809049}}.
\bibitem[{Adak et~al.(2013)Adak, Sert, Kalay, and Sari}]{Adak:2008gd}
\bibinfo{author}{M.~Adak}, \bibinfo{author}{O.~Sert},
  \bibinfo{author}{M.~Kalay}, \bibinfo{author}{M.~Sari},
\newblock \bibinfo{title}{{Symmetric Teleparallel Gravity: Some exact solutions
  and spinor couplings}},
\newblock \bibinfo{journal}{Int. J. Mod. Phys. A} \bibinfo{volume}{28}
  (\bibinfo{year}{2013}) \bibinfo{pages}{1350167}.
  \DOIprefix\doi{10.1142/S0217751X13501674}.
  \href{http://arxiv.org/abs/0810.2388}{{\tt arXiv:0810.2388}}.
\bibitem[{Adak(2018)}]{Adak:2018vzk}
\bibinfo{author}{M.~Adak},
\newblock \bibinfo{title}{{Gauge Approach to The Symmetric Teleparallel
  Gravity}},
\newblock \bibinfo{journal}{Int. J. Geom. Meth. Mod. Phys.}
  \bibinfo{volume}{15} (\bibinfo{year}{2018}) \bibinfo{pages}{1850198}.
  \DOIprefix\doi{10.1142/S0219887818501980}.
  \href{http://arxiv.org/abs/1809.01385}{{\tt arXiv:1809.01385}}.
\bibitem[{Jim\'enez et~al.(2019)Jim\'enez, Heisenberg, and
  Koivisto}]{BeltranJimenez:2019tjy}
\bibinfo{author}{J.~B. Jim\'enez}, \bibinfo{author}{L.~Heisenberg},
  \bibinfo{author}{T.~S. Koivisto},
\newblock \bibinfo{title}{{The Geometrical Trinity of Gravity}},
\newblock \bibinfo{journal}{Universe} \bibinfo{volume}{5}
  (\bibinfo{year}{2019}) \bibinfo{pages}{173}.
  \DOIprefix\doi{10.3390/universe5070173}.
  \href{http://arxiv.org/abs/1903.06830}{{\tt arXiv:1903.06830}}.
\bibitem[{Lazkoz et~al.(2019)Lazkoz, Lobo, Ortiz-Ba\~nos, and
  Salzano}]{Lazkoz:2019sjl}
\bibinfo{author}{R.~Lazkoz}, \bibinfo{author}{F.~S.~N. Lobo},
  \bibinfo{author}{M.~Ortiz-Ba\~nos}, \bibinfo{author}{V.~Salzano},
\newblock \bibinfo{title}{{Observational constraints of $f(Q)$ gravity}},
\newblock \bibinfo{journal}{Phys. Rev. D} \bibinfo{volume}{100}
  (\bibinfo{year}{2019}) \bibinfo{pages}{104027}.
  \DOIprefix\doi{10.1103/PhysRevD.100.104027}.
  \href{http://arxiv.org/abs/1907.13219}{{\tt arXiv:1907.13219}}.
\bibitem[{Ayuso et~al.(2021)Ayuso, Lazkoz, and Salzano}]{Ayuso:2020dcu}
\bibinfo{author}{I.~Ayuso}, \bibinfo{author}{R.~Lazkoz},
  \bibinfo{author}{V.~Salzano},
\newblock \bibinfo{title}{{Observational constraints on cosmological solutions
  of $f(Q)$ theories}},
\newblock \bibinfo{journal}{Phys. Rev. D} \bibinfo{volume}{103}
  (\bibinfo{year}{2021}) \bibinfo{pages}{063505}.
  \DOIprefix\doi{10.1103/PhysRevD.103.063505}.
  \href{http://arxiv.org/abs/2012.00046}{{\tt arXiv:2012.00046}}.
\bibitem[{Khyllep et~al.(2021)Khyllep, Paliathanasis, and
  Dutta}]{Khyllep:2021pcu}
\bibinfo{author}{W.~Khyllep}, \bibinfo{author}{A.~Paliathanasis},
  \bibinfo{author}{J.~Dutta},
\newblock \bibinfo{title}{{Cosmological solutions and growth index of matter
  perturbations in $f(Q)$ gravity}},
\newblock \bibinfo{journal}{Phys. Rev. D} \bibinfo{volume}{103}
  (\bibinfo{year}{2021}) \bibinfo{pages}{103521}.
  \DOIprefix\doi{10.1103/PhysRevD.103.103521}.
  \href{http://arxiv.org/abs/2103.08372}{{\tt arXiv:2103.08372}}.
\bibitem[{Anagnostopoulos et~al.(2021)Anagnostopoulos, Basilakos, and
  Saridakis}]{Anagnostopoulos:2021ydo}
\bibinfo{author}{F.~K. Anagnostopoulos}, \bibinfo{author}{S.~Basilakos},
  \bibinfo{author}{E.~N. Saridakis},
\newblock \bibinfo{title}{{First evidence that non-metricity f(Q) gravity can
  challenge $\Lambda$CDM}}  (\bibinfo{year}{2021}).
  \href{http://arxiv.org/abs/2104.15123}{{\tt arXiv:2104.15123}}.
\bibitem[{Frusciante(2021)}]{Frusciante:2021sio}
\bibinfo{author}{N.~Frusciante},
\newblock \bibinfo{title}{{Signatures of $f(Q)$-gravity in cosmology}},
\newblock \bibinfo{journal}{Phys. Rev. D} \bibinfo{volume}{103}
  (\bibinfo{year}{2021}) \bibinfo{pages}{044021}.
  \DOIprefix\doi{10.1103/PhysRevD.103.044021}.
  \href{http://arxiv.org/abs/2101.09242}{{\tt arXiv:2101.09242}}.
\bibitem[{Atayde and Frusciante(2021)}]{Atayde:2021ujc}
\bibinfo{author}{L.~Atayde}, \bibinfo{author}{N.~Frusciante},
\newblock \bibinfo{title}{{Can $f(Q)$ gravity challenge $\Lambda$CDM?}},
\newblock \bibinfo{journal}{Phys. Rev. D} \bibinfo{volume}{104}
  (\bibinfo{year}{2021}) \bibinfo{pages}{064052}.
  \DOIprefix\doi{10.1103/PhysRevD.104.064052}.
  \href{http://arxiv.org/abs/2108.10832}{{\tt arXiv:2108.10832}}.
\bibitem[{Dialektopoulos et~al.(2019)Dialektopoulos, Koivisto, and
  Capozziello}]{Dialektopoulos:2019mtr}
\bibinfo{author}{K.~F. Dialektopoulos}, \bibinfo{author}{T.~S. Koivisto},
  \bibinfo{author}{S.~Capozziello},
\newblock \bibinfo{title}{{Noether symmetries in Symmetric Teleparallel
  Cosmology}},
\newblock \bibinfo{journal}{Eur. Phys. J. C} \bibinfo{volume}{79}
  (\bibinfo{year}{2019}) \bibinfo{pages}{606}.
  \DOIprefix\doi{10.1140/epjc/s10052-019-7106-8}.
  \href{http://arxiv.org/abs/1905.09019}{{\tt arXiv:1905.09019}}.
\bibitem[{Beltr\'an~Jim\'enez et~al.(2020)Beltr\'an~Jim\'enez, Heisenberg,
  Koivisto, and Pekar}]{Jimenez:2019ovq}
\bibinfo{author}{J.~Beltr\'an~Jim\'enez}, \bibinfo{author}{L.~Heisenberg},
  \bibinfo{author}{T.~S. Koivisto}, \bibinfo{author}{S.~Pekar},
\newblock \bibinfo{title}{{Cosmology in $f(Q)$ geometry}},
\newblock \bibinfo{journal}{Phys. Rev. D} \bibinfo{volume}{101}
  (\bibinfo{year}{2020}) \bibinfo{pages}{103507}.
  \DOIprefix\doi{10.1103/PhysRevD.101.103507}.
  \href{http://arxiv.org/abs/1906.10027}{{\tt arXiv:1906.10027}}.
\bibitem[{Bajardi et~al.(2020)Bajardi, Vernieri, and
  Capozziello}]{Bajardi:2020fxh}
\bibinfo{author}{F.~Bajardi}, \bibinfo{author}{D.~Vernieri},
  \bibinfo{author}{S.~Capozziello},
\newblock \bibinfo{title}{{Bouncing Cosmology in f(Q) Symmetric Teleparallel
  Gravity}},
\newblock \bibinfo{journal}{Eur. Phys. J. Plus} \bibinfo{volume}{135}
  (\bibinfo{year}{2020}) \bibinfo{pages}{912}.
  \DOIprefix\doi{10.1140/epjp/s13360-020-00918-3}.
  \href{http://arxiv.org/abs/2011.01248}{{\tt arXiv:2011.01248}}.
\bibitem[{Flathmann and Hohmann(2021)}]{Flathmann:2020zyj}
\bibinfo{author}{K.~Flathmann}, \bibinfo{author}{M.~Hohmann},
\newblock \bibinfo{title}{{Post-Newtonian limit of generalized symmetric
  teleparallel gravity}},
\newblock \bibinfo{journal}{Phys. Rev. D} \bibinfo{volume}{103}
  (\bibinfo{year}{2021}) \bibinfo{pages}{044030}.
  \DOIprefix\doi{10.1103/PhysRevD.103.044030}.
  \href{http://arxiv.org/abs/2012.12875}{{\tt arXiv:2012.12875}}.
\bibitem[{D'Ambrosio et~al.(2020)D'Ambrosio, Garg, and
  Heisenberg}]{DAmbrosio:2020nev}
\bibinfo{author}{F.~D'Ambrosio}, \bibinfo{author}{M.~Garg},
  \bibinfo{author}{L.~Heisenberg},
\newblock \bibinfo{title}{{Non-linear extension of non-metricity scalar for
  MOND}},
\newblock \bibinfo{journal}{Phys. Lett. B} \bibinfo{volume}{811}
  (\bibinfo{year}{2020}) \bibinfo{pages}{135970}.
  \DOIprefix\doi{10.1016/j.physletb.2020.135970}.
  \href{http://arxiv.org/abs/2004.00888}{{\tt arXiv:2004.00888}}.
\bibitem[{Solanki et~al.(2021)Solanki, Pacif, Parida, and
  Sahoo}]{Solanki:2021qni}
\bibinfo{author}{R.~Solanki}, \bibinfo{author}{S.~K.~J. Pacif},
  \bibinfo{author}{A.~Parida}, \bibinfo{author}{P.~K. Sahoo},
\newblock \bibinfo{title}{{Cosmic acceleration with bulk viscosity in modified
  f(Q) gravity}},
\newblock \bibinfo{journal}{Phys. Dark Univ.} \bibinfo{volume}{32}
  (\bibinfo{year}{2021}) \bibinfo{pages}{100820}.
  \DOIprefix\doi{10.1016/j.dark.2021.100820}.
  \href{http://arxiv.org/abs/2105.00876}{{\tt arXiv:2105.00876}}.
\bibitem[{Zhao(2021)}]{Zhao:2021zab}
\bibinfo{author}{D.~Zhao},
\newblock \bibinfo{title}{{Covariant formulation of f(Q) theory}}
  (\bibinfo{year}{2021}). \href{http://arxiv.org/abs/2104.02483}{{\tt
  arXiv:2104.02483}}.
\bibitem[{B\"ohmer and Jensko(2021)}]{Bohmer:2021eoo}
\bibinfo{author}{C.~G. B\"ohmer}, \bibinfo{author}{E.~Jensko},
\newblock \bibinfo{title}{{Modified gravity: a unified approach}}
  (\bibinfo{year}{2021}). \href{http://arxiv.org/abs/2103.15906}{{\tt
  arXiv:2103.15906}}.
\bibitem[{Song et~al.(2007)Song, Hu, and Sawicki}]{Song:0610532}
\bibinfo{author}{Y.-S. Song}, \bibinfo{author}{W.~Hu},
  \bibinfo{author}{I.~Sawicki},
\newblock \bibinfo{title}{{The Large Scale Structure of f(R) Gravity}},
\newblock \bibinfo{journal}{Phys. Rev. D} \bibinfo{volume}{75}
  (\bibinfo{year}{2007}) \bibinfo{pages}{044004}.
  \DOIprefix\doi{10.1103/PhysRevD.75.044004}.
  \href{http://arxiv.org/abs/astro-ph/0610532}{{\tt arXiv:astro-ph/0610532}}.
\bibitem[{Pogosian and Silvestri(2008)}]{Pogosian:0709}
\bibinfo{author}{L.~Pogosian}, \bibinfo{author}{A.~Silvestri},
\newblock \bibinfo{title}{{The pattern of growth in viable f(R) cosmologies}},
\newblock \bibinfo{journal}{Phys. Rev. D} \bibinfo{volume}{77}
  (\bibinfo{year}{2008}) \bibinfo{pages}{023503}.
  \DOIprefix\doi{10.1103/PhysRevD.77.023503}.
  \href{http://arxiv.org/abs/0709.0296}{{\tt arXiv:0709.0296}},
  \bibinfo{note}{[Erratum: Phys.Rev.D 81, 049901 (2010)]}.
\bibitem[{Aldrovandi and Pereira(2013)}]{Aldrovandi:2013wha}
\bibinfo{author}{R.~Aldrovandi}, \bibinfo{author}{J.~G. Pereira},
  \bibinfo{title}{{Teleparallel Gravity}: {An Introduction}},
  \bibinfo{publisher}{Springer}, \bibinfo{year}{2013}.
  \DOIprefix\doi{10.1007/978-94-007-5143-9}.
\bibitem[{Beltr\'an~Jim\'enez et~al.(2020)Beltr\'an~Jim\'enez, Heisenberg,
  Koivisto, and Pekar}]{BeltranJimenez:2019tme}
\bibinfo{author}{J.~Beltr\'an~Jim\'enez}, \bibinfo{author}{L.~Heisenberg},
  \bibinfo{author}{T.~S. Koivisto}, \bibinfo{author}{S.~Pekar},
\newblock \bibinfo{title}{{Cosmology in $f(Q)$ geometry}},
\newblock \bibinfo{journal}{Phys. Rev. D} \bibinfo{volume}{101}
  (\bibinfo{year}{2020}) \bibinfo{pages}{103507}.
  \DOIprefix\doi{10.1103/PhysRevD.101.103507}.
  \href{http://arxiv.org/abs/1906.10027}{{\tt arXiv:1906.10027}}.
\bibitem[{De~Felice et~al.(2012)De~Felice, Nesseris, and
  Tsujikawa}]{DeFelice:1203}
\bibinfo{author}{A.~De~Felice}, \bibinfo{author}{S.~Nesseris},
  \bibinfo{author}{S.~Tsujikawa},
\newblock \bibinfo{title}{{Observational constraints on dark energy with a fast
  varying equation of state}},
\newblock \bibinfo{journal}{JCAP} \bibinfo{volume}{05} (\bibinfo{year}{2012})
  \bibinfo{pages}{029}. \DOIprefix\doi{10.1088/1475-7516/2012/05/029}.
  \href{http://arxiv.org/abs/1203.6760}{{\tt arXiv:1203.6760}}.
\bibitem[{Aghanim et~al.(2020)}]{Planck:2018vyg}
\bibinfo{author}{N.~Aghanim}, et~al. (\bibinfo{collaboration}{Planck}),
\newblock \bibinfo{title}{{Planck 2018 results. VI. Cosmological parameters}},
\newblock \bibinfo{journal}{Astron. Astrophys.} \bibinfo{volume}{641}
  (\bibinfo{year}{2020}) \bibinfo{pages}{A6}.
  \DOIprefix\doi{10.1051/0004-6361/201833910}.
  \href{http://arxiv.org/abs/1807.06209}{{\tt arXiv:1807.06209}}.
\bibitem[{Amendola et~al.(2008)Amendola, Kunz, and Sapone}]{Amendola:2007rr}
\bibinfo{author}{L.~Amendola}, \bibinfo{author}{M.~Kunz},
  \bibinfo{author}{D.~Sapone},
\newblock \bibinfo{title}{{Measuring the dark side (with weak lensing)}},
\newblock \bibinfo{journal}{JCAP} \bibinfo{volume}{04} (\bibinfo{year}{2008})
  \bibinfo{pages}{013}. \DOIprefix\doi{10.1088/1475-7516/2008/04/013}.
  \href{http://arxiv.org/abs/0704.2421}{{\tt arXiv:0704.2421}}.
\bibitem[{Bean and Tangmatitham(2010)}]{PhysRevD.81.083534}
\bibinfo{author}{R.~Bean}, \bibinfo{author}{M.~Tangmatitham},
\newblock \bibinfo{title}{Current constraints on the cosmic growth history},
\newblock \bibinfo{journal}{Phys. Rev. D} \bibinfo{volume}{81}
  (\bibinfo{year}{2010}) \bibinfo{pages}{083534}. \URLprefix
  \url{https://link.aps.org/doi/10.1103/PhysRevD.81.083534}.
  \DOIprefix\doi{10.1103/PhysRevD.81.083534}.
\bibitem[{Silvestri et~al.(2013)Silvestri, Pogosian, and
  Buniy}]{Silvestri:2013ne}
\bibinfo{author}{A.~Silvestri}, \bibinfo{author}{L.~Pogosian},
  \bibinfo{author}{R.~V. Buniy},
\newblock \bibinfo{title}{{Practical approach to cosmological perturbations in
  modified gravity}},
\newblock \bibinfo{journal}{Phys. Rev. D} \bibinfo{volume}{87}
  (\bibinfo{year}{2013}) \bibinfo{pages}{104015}.
  \DOIprefix\doi{10.1103/PhysRevD.87.104015}.
  \href{http://arxiv.org/abs/1302.1193}{{\tt arXiv:1302.1193}}.
\bibitem[{{Pogosian} et~al.(2010){Pogosian}, {Silvestri}, {Koyama}, and
  {Zhao}}]{2010PhRvD..81j4023P}
\bibinfo{author}{L.~{Pogosian}}, \bibinfo{author}{A.~{Silvestri}},
  \bibinfo{author}{K.~{Koyama}}, \bibinfo{author}{G.-B. {Zhao}},
\newblock \bibinfo{title}{{How to optimally parametrize deviations from general
  relativity in the evolution of cosmological perturbations}},
\newblock \bibinfo{journal}{Phys. Rev. D} \bibinfo{volume}{81}
  (\bibinfo{year}{2010}) \bibinfo{pages}{104023}.
  \DOIprefix\doi{10.1103/PhysRevD.81.104023}.
  \href{http://arxiv.org/abs/1002.2382}{{\tt arXiv:1002.2382}}.
\bibitem[{Amendola et~al.(2020)Amendola, Bettoni, Pinho, and
  Casas}]{Amendola:2019laa}
\bibinfo{author}{L.~Amendola}, \bibinfo{author}{D.~Bettoni},
  \bibinfo{author}{A.~M. Pinho}, \bibinfo{author}{S.~Casas},
\newblock \bibinfo{title}{{Measuring gravity at cosmological scales}},
\newblock \bibinfo{journal}{Universe} \bibinfo{volume}{6}
  (\bibinfo{year}{2020}) \bibinfo{pages}{20}.
  \DOIprefix\doi{10.3390/universe6020020}.
  \href{http://arxiv.org/abs/1902.06978}{{\tt arXiv:1902.06978}}.
\bibitem[{Sagredo et~al.(2018)Sagredo, Nesseris, and Sapone}]{Sagredo:2018ahx}
\bibinfo{author}{B.~Sagredo}, \bibinfo{author}{S.~Nesseris},
  \bibinfo{author}{D.~Sapone},
\newblock \bibinfo{title}{{Internal Robustness of Growth Rate data}},
\newblock \bibinfo{journal}{Phys. Rev. D} \bibinfo{volume}{98}
  (\bibinfo{year}{2018}) \bibinfo{pages}{083543}.
  \DOIprefix\doi{10.1103/PhysRevD.98.083543}.
  \href{http://arxiv.org/abs/1806.10822}{{\tt arXiv:1806.10822}}.
\bibitem[{Renk et~al.(2017)Renk, Zumalac\'arregui, Montanari, and
  Barreira}]{Renk:2017rzu}
\bibinfo{author}{J.~Renk}, \bibinfo{author}{M.~Zumalac\'arregui},
  \bibinfo{author}{F.~Montanari}, \bibinfo{author}{A.~Barreira},
\newblock \bibinfo{title}{{Galileon gravity in light of ISW, CMB, BAO and H$_0$
  data}},
\newblock \bibinfo{journal}{JCAP} \bibinfo{volume}{10} (\bibinfo{year}{2017})
  \bibinfo{pages}{020}. \DOIprefix\doi{10.1088/1475-7516/2017/10/020}.
  \href{http://arxiv.org/abs/1707.02263}{{\tt arXiv:1707.02263}}.
\bibitem[{Song et~al.(2007)Song, Hu, and Sawicki}]{Song:2006ej}
\bibinfo{author}{Y.-S. Song}, \bibinfo{author}{W.~Hu},
  \bibinfo{author}{I.~Sawicki},
\newblock \bibinfo{title}{{The Large Scale Structure of f(R) Gravity}},
\newblock \bibinfo{journal}{Phys. Rev. D} \bibinfo{volume}{75}
  (\bibinfo{year}{2007}) \bibinfo{pages}{044004}.
  \DOIprefix\doi{10.1103/PhysRevD.75.044004}.
  \href{http://arxiv.org/abs/astro-ph/0610532}{{\tt arXiv:astro-ph/0610532}}.
\bibitem[{Barreira et~al.(2012)Barreira, Li, Baugh, and
  Pascoli}]{Barreira:2012kk}
\bibinfo{author}{A.~Barreira}, \bibinfo{author}{B.~Li}, \bibinfo{author}{C.~M.
  Baugh}, \bibinfo{author}{S.~Pascoli},
\newblock \bibinfo{title}{{Linear perturbations in Galileon gravity models}},
\newblock \bibinfo{journal}{Phys. Rev. D} \bibinfo{volume}{86}
  (\bibinfo{year}{2012}) \bibinfo{pages}{124016}.
  \DOIprefix\doi{10.1103/PhysRevD.86.124016}.
  \href{http://arxiv.org/abs/1208.0600}{{\tt arXiv:1208.0600}}.
\bibitem[{Giacomello et~al.(2019)Giacomello, De~Felice, and
  Ansoldi}]{Giacomello:2018jfi}
\bibinfo{author}{F.~Giacomello}, \bibinfo{author}{A.~De~Felice},
  \bibinfo{author}{S.~Ansoldi},
\newblock \bibinfo{title}{{Bounds from ISW-galaxy cross-correlations on
  generalized covariant Galileon models}},
\newblock \bibinfo{journal}{JCAP} \bibinfo{volume}{03} (\bibinfo{year}{2019})
  \bibinfo{pages}{038}. \DOIprefix\doi{10.1088/1475-7516/2019/03/038}.
  \href{http://arxiv.org/abs/1811.10885}{{\tt arXiv:1811.10885}}.
\bibitem[{Hang et~al.(2021)Hang, Alam, Cai, and Peacock}]{Hang:2021kfx}
\bibinfo{author}{Q.~Hang}, \bibinfo{author}{S.~Alam}, \bibinfo{author}{Y.-C.
  Cai}, \bibinfo{author}{J.~A. Peacock},
\newblock \bibinfo{title}{{Stacked CMB lensing and ISW signals around
  superstructures in the DESI Legacy Survey}},
\newblock \bibinfo{journal}{Mon. Not. Roy. Astron. Soc.} \bibinfo{volume}{507}
  (\bibinfo{year}{2021}) \bibinfo{pages}{510--523}.
  \DOIprefix\doi{10.1093/mnras/stab2184}.
  \href{http://arxiv.org/abs/2105.11936}{{\tt arXiv:2105.11936}}.
\bibitem[{Kable et~al.(2021)Kable, Benevento, Frusciante, De~Felice, and
  Tsujikawa}]{Kable:2021yws}
\bibinfo{author}{J.~A. Kable}, \bibinfo{author}{G.~Benevento},
  \bibinfo{author}{N.~Frusciante}, \bibinfo{author}{A.~De~Felice},
  \bibinfo{author}{S.~Tsujikawa},
\newblock \bibinfo{title}{{Probing Modified Gravity with Integrated Sachs-Wolfe
  CMB and Galaxy Cross-correlations}}  (\bibinfo{year}{2021}).
  \href{http://arxiv.org/abs/2111.10432}{{\tt arXiv:2111.10432}}.
\bibitem[{Kimura et~al.(2012)Kimura, Kobayashi, and Yamamoto}]{Kimura:2011td}
\bibinfo{author}{R.~Kimura}, \bibinfo{author}{T.~Kobayashi},
  \bibinfo{author}{K.~Yamamoto},
\newblock \bibinfo{title}{{Observational Constraints on Kinetic Gravity
  Braiding from the Integrated Sachs-Wolfe Effect}},
\newblock \bibinfo{journal}{Phys. Rev. D} \bibinfo{volume}{85}
  (\bibinfo{year}{2012}) \bibinfo{pages}{123503}.
  \DOIprefix\doi{10.1103/PhysRevD.85.123503}.
  \href{http://arxiv.org/abs/1110.3598}{{\tt arXiv:1110.3598}}.
\bibitem[{Nakamura et~al.(2019)Nakamura, De~Felice, Kase, and
  Tsujikawa}]{Nakamura:1811}
\bibinfo{author}{S.~Nakamura}, \bibinfo{author}{A.~De~Felice},
  \bibinfo{author}{R.~Kase}, \bibinfo{author}{S.~Tsujikawa},
\newblock \bibinfo{title}{{Constraints on massive vector dark energy models
  from integrated Sachs-Wolfe-galaxy cross-correlations}},
\newblock \bibinfo{journal}{Phys. Rev. D} \bibinfo{volume}{99}
  (\bibinfo{year}{2019}) \bibinfo{pages}{063533}.
  \DOIprefix\doi{10.1103/PhysRevD.99.063533}.
  \href{http://arxiv.org/abs/1811.07541}{{\tt arXiv:1811.07541}}.
\bibitem[{De~Felice et~al.(2011)De~Felice, Kobayashi, and
  Tsujikawa}]{DeFelice:2011hq}
\bibinfo{author}{A.~De~Felice}, \bibinfo{author}{T.~Kobayashi},
  \bibinfo{author}{S.~Tsujikawa},
\newblock \bibinfo{title}{{Effective gravitational couplings for cosmological
  perturbations in the most general scalar-tensor theories with second-order
  field equations}},
\newblock \bibinfo{journal}{Phys. Lett. B} \bibinfo{volume}{706}
  (\bibinfo{year}{2011}) \bibinfo{pages}{123--133}.
  \DOIprefix\doi{10.1016/j.physletb.2011.11.028}.
  \href{http://arxiv.org/abs/1108.4242}{{\tt arXiv:1108.4242}}.

\end{thebibliography}

\end{document}